\documentclass[12pt]{article}

\newif\ifdraft
\draftfalse

\usepackage[T1]{fontenc}
\usepackage[utf8]{inputenc}

\usepackage{mathptmx}

\usepackage[margin=1.2in]{geometry}
\usepackage{microtype}

\usepackage{amsmath,amssymb,amsthm}
\usepackage{bm}
\usepackage{graphicx}
\usepackage{booktabs}
\usepackage{multirow}
\usepackage{array}
\usepackage{caption}
\usepackage{subcaption}
\usepackage{xcolor}
\usepackage[numbers,sort&compress]{natbib}
\usepackage[hidelinks]{hyperref}
\usepackage[capitalise,nameinlink]{cleveref}

\usepackage{lineno}

\theoremstyle{plain}

\theoremstyle{definition}

\theoremstyle{remark}

\crefname{theorem}{Theorem}{Theorems}
\crefname{proposition}{Proposition}{Propositions}
\crefname{corollary}{Corollary}{Corollaries}
\crefname{definition}{Definition}{Definitions}

\IfFileExists{algorithm.sty}{%
  \usepackage{algorithm}%
  \IfFileExists{algpseudocode.sty}{\usepackage{algpseudocode}%
    }{}%
}{%
}

\newcommand{\Geff}{G_{\mathrm{eff}}}
\newcommand{\Pshare}{P^{\mathrm{share}}}
\newcommand{\Rt}{\widetilde{R}}          
\newcommand{\rt}{\widetilde{r}}          
\newcommand{\gvirt}{g_{\mathrm{virtual}}}

\newcommand{\gstar}{g^{*}}
\newcommand{\etacut}{\eta_{\mathrm{cut}}}
\newcommand{\kmat}{k_{\mathrm{m}}}
\newcommand{\aK}{a_{K}}                  
\newcommand{\Rthr}{R_{\mathrm{thr}}}     
\newcommand{\WK}{\,\mathrm{W/K}}
\newcommand{\WmK}{\,\mathrm{W/(m{\cdot}K)}}
\newcommand{\mKW}{\,\mathrm{m^{2}{\cdot}K/W}}

\newcommand{\um}{\,\mu\mathrm{m}}
\newcommand{\dB}{d_{\mathrm{B}}}         

\newcommand{\thcrit}{\theta^{*}}                   

\usepackage{etoolbox}
\apptocmd{\thebibliography}{\raggedright\sloppy\setlength{\emergencystretch}{3em}}{}{}

\title{The conductance filtration: sink-relative persistence and exact edge
  sensitivity for heat transport in particle-filled composites}
\author{Daisuke Yasufuku\thanks{Corresponding author:
    \texttt{yasufuku@umap-corp.com}}\\
  U-MAP Co., Ltd., Nagoya, Japan\\
  ORCID: \texttt{https://orcid.org/0009-0008-6555-5724}}
\date{\today}

\begin{document}
\ifdraft\linenumbers\fi
\maketitle

\begin{abstract}
We formulate the thermal conductance $g_e$ of a particle pair as three
branch types -- contact, near-contact and bond -- within a single
construction, and we introduce the filtration that takes this $g_e$ as its
axis (the conductance filtration) together with persistence taken relative
to the sink boundary. We then show that the dissipation share is exactly
the logarithmic sensitivity of the effective conductance, and that a finite
change of a single edge is predicted exactly in closed form. The allocation
of the dissipation among branch types, and the idealisation ratio of each
mechanism (interface, matrix and particle), are thereby obtained as
diagnostic quantities on one and the same graph representation.

\end{abstract}

\section{Introduction}\label{sec:1}

The heat generated per unit area of electronic devices keeps rising as
performance and packaging density increase. In data centres the power
density per rack of computers has risen by an order of magnitude over the
past few years~\citep{iea2026}. The growth in heat generation drives
stronger cooling, and the transition from air cooling to liquid cooling is
under way. The lower the thermal resistance on the cooling side, the larger
the share of the total thermal resistance of the system that the interfaces
take. In a test vehicle reproducing the geometry and the mechanical
constraints of a representative microprocessor package, the thermal
interface was reported to account for more than 60\% of the total thermal
resistance~\citep{linderman2007}. The material that fills this interface is
a thermal interface material (TIM), a composite in which a
thermally conductive filler is loaded into a matrix at a high packing
fraction.

High loading as a design strategy follows from the thermal conductivity of
the matrix, which remains at $0.1$--$0.4\WmK$. Conduction paths through the
matrix have to be built out of filler, and loadings reaching a volume
fraction of 78\% are used in practice~\citep{linderman2007}. There are
limits, however. First, cases are reported in which an increase in the
nominal thermal conductivity of the filler does not translate into a
reduction of the effective thermal resistance: commercial thermal greases
of nominal $8.5\WmK$ and $14.2\WmK$ both remain at $3$--$6\,\mathrm{mm^{2}
{\cdot}K/W}$ in the same measurement
system~\citep{cheng2025}. Second, an increase in the packing fraction
raises the bulk thermal conductivity exponentially, but it raises the
viscosity exponentially as well. As a result the thermal conductivity
achieved by a particulate composite remains about an order of magnitude
below the thermal conductivity of the particles
themselves~\citep{linderman2007}.

Behind this situation lies a structural problem: the physical process that
lies between the design variables and the quantity that is evaluated is not
observed. The designer can act on four variables -- the packing fraction
$\phi$, the particle size distribution, the filler species and the surface
treatment. What the evaluation returns, in contrast, is a single quantity,
the effective thermal conductivity. Between the two there is a process: the
packed structure fixes the contact and near-contact geometry of the
particle pairs, that geometry together with the material properties fixes
the thermal conductance of each pair, those conductances form a network,
and transport on that network yields the effective thermal conductivity.
Each stage of this process is computable in isolation, yet a standard
evaluation returns only the outcome of passing through all of them.
Consequently, when the measured conductivity falls short of the target, one
cannot decide which of the four variables to change, or by how much. For
this reason, elucidating the internal heat transfer mechanisms has been put
forward as a direction for improving the thermal conductivity of composites
once they have been fabricated~\citep{gu2021}.

This paper makes each stage of that process available as a computable
quantity. We construct a graph representation whose edges are the thermal
conductances of particle pairs, and on it we provide the identification of
the elements that are limiting and the prediction of the response to a
specified change.

Previous modelling of the thermal conductivity of particulate composites
falls broadly into two lines of work. The first is the traditional approach
that yields effective properties, or an indicator of them, from the
constituent properties and the structure. Effective medium theory (EMT)
yields the effective value directly from the packing fraction, the
conductivities of the constituents and the interfacial thermal
resistance~\citep{nan1997}. The discrete network approximation maps the
structure onto a network of nodes and edges and solves the transport
equation on it; for high-contrast, densely packed composites it carries an
asymptotic justification with a priori error estimates independent of the
number of particles~\citep{berlyand2012}. For random assemblies of spheres
there is numerical evidence of agreement with the finite element method to
within 4\% on average~\citep{birkholz2019}. A mechanistic model aimed at
the dense regime assembles the branch thermal resistances over a
body-centred cubic unit cell and yields accurate predictions across a wide
range of packing fractions~\citep{he2026}. There are also attempts to
extract features from the network using graph theory, in which quantities
obtained from a directed weighted thermal network are shown to correlate
with the anisotropic effective thermal conductivity~\citep{fei2022}. What
these have in common is that they return one quantity, or a few, for one
structure; none of them is a framework that returns, for an arbitrary
packed structure, a breakdown of the mechanisms that are limiting that
value. The mechanistic model does treat the branch resistances explicitly,
but its unit cell is an ideal lattice, so it does not accept an arbitrary
packed structure as input.

The second line of work is the recent approach that yields a
multi-dimensional descriptor from the structure itself, centred on
persistent homology~\citep{hiraoka2016,robins2016,minamitani2022,%
miyazaki2025}. It differs from the first in that it does not contract the
structure to a single number but retains information across scales. The
axis of all of these filtrations is a geometric quantity, however, and the
only instance of a physical quantity as the axis is the contact force in
granular matter~\citep{kramar2013}.

The position of this paper is the connection between these two lines. We
use the physical quantity the first line handles, namely the thermal
conductance $g_e$ of each particle pair, as the axis of the filtration,
which is the method of the second. The construction of a superlevel set
filtration on edge weights is itself standard~\citep{aktas2019}; what we
introduce is the choice of the function and the relativisation with respect
to the sink boundary.

\begin{figure}[tbp]
  \centering
  \includegraphics[width=\linewidth]{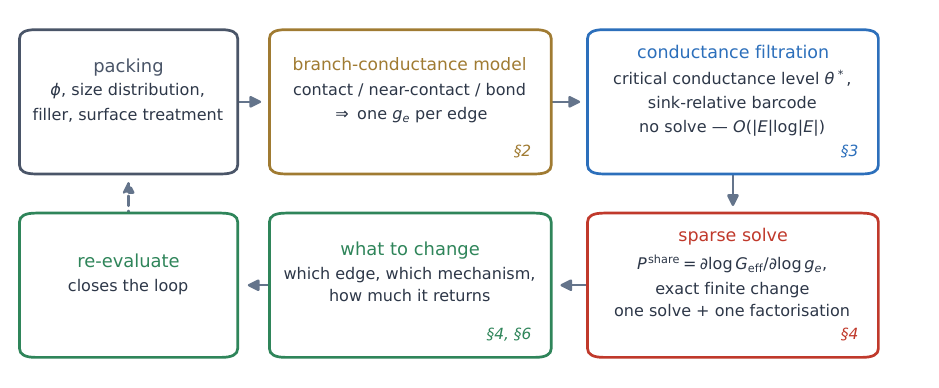}
  \caption{The design loop closes on a single graph representation. The
  four design variables of \cref{sec:1} enter on the left; the branch
  conductances, the conductance filtration and the dissipation share are
  computed on one graph; and the diagnosis returns to the design
  variables.}\label{fig:1}
\end{figure}

We make four contributions. First, we formulate a branch-conductance model
that treats the contact, near-contact and bond branches within a single
construction, put each term in correspondence with existing theory, and
state the conditions of validity explicitly (\cref{sec:2}). Second, we
introduce the conductance filtration and persistence relative to the sink
boundary (\cref{sec:3}). Third, we show that the dissipation share is
exactly the logarithmic sensitivity of the effective conductance, and that
a finite change of a single edge is predicted exactly in closed form
(\cref{sec:4}). Fourth, we give algorithms that carry all of this out with
one sparse solve and one factorisation, implement them as an accompanying
software package that reproduces every figure of this paper
(\cref{sec:5}), and present a numerical demonstration on eight conditions
formed from two particle sizes, two fillers and the surface treatment
(\cref{sec:6}). Taken together, the design loop -- from generating the
structure through identifying the limiting mechanism, evaluating a change
and re-evaluating -- closes on a single graph representation. This
procedure is shown in \cref{fig:1}.

The paper is organised as follows: \cref{sec:2} the branch-conductance
model, \cref{sec:3} the filtration, \cref{sec:4} the sensitivity,
\cref{sec:5} the implementation, \cref{sec:6} the numerical demonstration,
\cref{sec:7} the limits, and \cref{sec:8} the conclusion.

\section{The branch-conductance model}\label{sec:2}

\subsection{The branch-conductance model}\label{sec:2.1}

In this section we give the rule that fixes the thermal conductance $g_e$
of each particle pair from the arrangement of the particles and the
properties of the materials. We call it the branch-conductance model.

We map the packed structure onto a weighted graph $G=(V,E,g)$ whose
vertices are the particles and whose edges are the particle pairs that can
exchange heat. Vertex $i$ is a sphere of radius $r_i$ and thermal
conductivity $\kappa_i$, and the thermal conductivity of the matrix is
$\kmat$. We assume that the interior of a particle is isothermal and assign
a single temperature $u_i$ to the vertex; we refer to this treatment as an
isothermal node. We take a region of fixed thickness from each of two
opposing faces of the specimen, call it a boundary slab, and take the sets
of particles whose centres lie in them as the source $\mathrm{src}$ and the
sink $\mathrm{snk}$, imposing $u|_{\mathrm{src}}=1$ and
$u|_{\mathrm{snk}}=0$. The total heat flow from source to sink under these
conditions is the effective conductance $\Geff$ that this paper evaluates,
which converts to the effective thermal conductivity as
$k_{\mathrm{eff}}=\Geff\,L/A$, where $L$ is the specimen length along the
heat flow and $A$ is the cross-sectional area. We use the reduced radius of
a pair,
\begin{equation}
  \rt=\frac{r_ir_j}{r_i+r_j}.
  \label{eq:rtilde}
\end{equation}

We write $h$ for the surface separation, with $h>0$ a separation and $h<0$
an overlap, and we call $\delta=-h>0$ the overlap depth. We write $R_{s,i}$
for the interfacial thermal resistance per unit area between particle $i$
and the matrix, and $R_{c,i}$ for the corresponding quantity on the side of
particle $i$ at a face where two particles touch directly. The interfaces
of both particles enter the heat path in series. On a near-contact branch,
$R_{s,i}$ enters as a length added to the effective gap,
$a_{K,i}=R_{s,i}k_{\mathrm{gap}}$, in the pair sum $a_{K,i}+a_{K,j}$; on a
contact branch the interfacial resistance of the contact face is
$(R_{c,i}+R_{c,j})/(\pi a^2)$ (\cref{sec:2.3.1}). The quantity $a_{K,i}$
has the dimension of a length and is called the Kapitza radius in the
composites literature~\citep{nan1997}. When the value is common to all
particles we drop the index and write $R_s$, $R_c$ and $\aK$. As for the
conductivity on the gap side, $k_{\mathrm{gap}}$, this paper expresses the
surface treatment as a change of the interfacial resistance and does not
introduce a layer of finite thickness ($t_{\mathrm{ip}}=0$), so
$k_{\mathrm{gap}}=\kmat$. The case with a layer is discussed in
\cref{sec:7.2}.

Which paths run through a particle pair is decided by two conditions:
whether the surfaces of the two particles are in contact, and whether the
two particles belong to the same continuous solid. These sort the pairs
into three kinds. We call a pair separated only by a thin film of matrix a
\emph{near-contact branch} (the subscript prox below stands for
proximity), a pair whose solids touch a \emph{contact branch}, and a pair
belonging to the same continuous solid a \emph{bond branch}.

\begin{figure}[htbp]
  \centering
  \includegraphics[width=\linewidth]{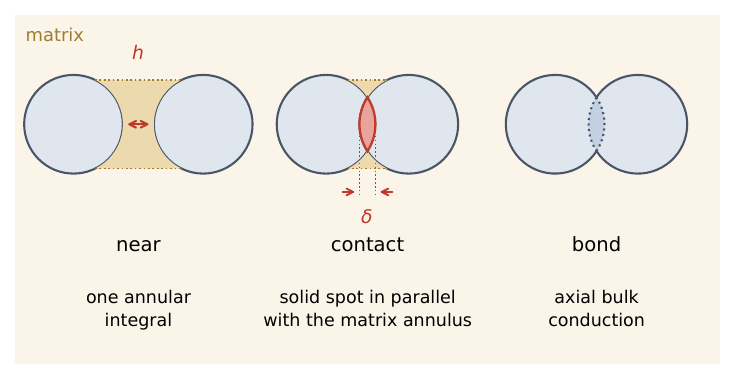}
  \caption{The three branch types. A pair separated only by a thin film of
  matrix (near-contact), a pair whose solids touch (contact), and a pair
  inside the same continuous solid (bond).}\label{fig:2}
\end{figure}

\begin{figure}[htbp]
  \centering
  \includegraphics[width=\linewidth]{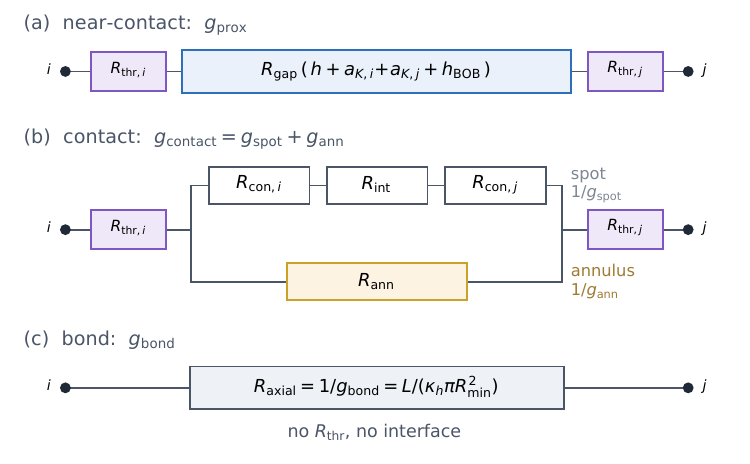}
  \caption{The thermal resistance network of each branch type, in the same
  order as the text. The through-particle resistance $\Rthr$ is in series
  with the whole edge, not inside the contact spot.}\label{fig:3}
\end{figure}

Earlier models either collect these into a single series chain of
resistances or adopt only one of them~\citep{birkholz2019,he2026}. The
three geometries are shown in \cref{fig:2} and the corresponding thermal
resistance networks in \cref{fig:3}. The expressions in this paper are
written in terms of the conductance $g$, and each element of the network
diagrams is its reciprocal thermal resistance $R=1/g$.

The matrix annulus of the contact branch follows from the same integral as
the near-contact branch, so we treat the near-contact branch first. The
derivations are given in \cref{app:A}; here we state the final form and the
physical meaning of each term.

\subsection{The near-contact branch}\label{sec:2.2}

On a pair with $h>0$ the solids are not in contact, so the heat passes
through the thin film of matrix alone. Both the interfacial resistance and
the saturation due to diffusion inside the particle appear as lengths added
to the effective gap:
\begin{equation}
  g_{\mathrm{prox}}=2\pi k_{\mathrm{gap}}\,\rt\,
  \ln\!\Big(1+\frac{\rt}{h+a_{K,i}+a_{K,j}+h_{\mathrm{BOB}}}\Big),
  \label{eq:gprox}
\end{equation}
and in the network of \cref{fig:3}(a) this edge is represented by the
single element $R_{\mathrm{gap}}=1/g_{\mathrm{prox}}$. We now give the
origin of each term.

\subsubsection{The annular integral and the interfacial term
  \texorpdfstring{$a_{K,i}+a_{K,j}$}{a\_Ki+a\_Kj}}\label{sec:2.2.1}

Under the thin-film (Derjaguin) approximation $h\ll\rt$, the gap between
the particles at a distance $r$ from the axis of symmetry is
$h(r)=h+r^2/(2\rt)$. Regarding each flux tube of radius $r$ as passing in
turn through interface, matrix and interface, and combining these in
parallel in the radial direction, one obtains the expression at the head of
this subsection with $h_{\mathrm{BOB}}$ removed (the derivation is in
\cref{app:A.2}). The interfacial resistance appears as the length
$h\to h+a_{K,i}+a_{K,j}$ rather than as an independent series term because,
within each flux tube combined in parallel in the radial direction,
$R_{s,i}+R_{s,j}$ enters at the same place as $h(r)/k_{\mathrm{gap}}$ (the
derivation is in \cref{app:A.2}). The prefactor $2\pi k_{\mathrm{gap}}\rt$
agrees with the prefactor in the expression Batchelor and O'Brien give for
the heat flux near a point of contact~\citep{batchelor1977}.

\subsubsection{The saturation term \texorpdfstring{$h_{\mathrm{BOB}}$}{h\_BOB}}\label{sec:2.2.2}

What sets the level of the saturation is the ratio of the conductivities of
the particle and the matrix, $\alpha=\kappa_h/\kmat$, where
$\kappa_h=2/(\kappa_i^{-1}+\kappa_j^{-1})$ is the harmonic mean over the
pair. Our model treats the resistance to diffusion inside the particle, like
the interface, as a length $h_{\mathrm{BOB}}=\rt/(\alpha^2-1)$ added to the
gap. This form is fixed uniquely by requiring that at $h=0$ and $\aK=0$ it
agree exactly with the point-contact saturation value
$4\pi k_{\mathrm{gap}}\rt\ln\alpha$ of Batchelor and
O'Brien~\citep{batchelor1977}.

\subsection{The contact branch}\label{sec:2.3}

When particles touch, a path directly connecting the solids is added in
parallel with the path through the matrix that remains around it, so that
\begin{equation}
  g_{\mathrm{contact}}=g_{\mathrm{spot}}+g_{\mathrm{ann}},
  \label{eq:gcontact}
\end{equation}
where $g_{\mathrm{spot}}$ is the path through the solid contact face and
$g_{\mathrm{ann}}$ the path through the matrix annulus that remains around
it. In the network of \cref{fig:3}(b), the matrix annulus
$R_{\mathrm{ann}}=1/g_{\mathrm{ann}}$ is in parallel with the series
combination $1/g_{\mathrm{spot}}=R_{\mathrm{con},i}+R_{\mathrm{int}}
+R_{\mathrm{con},j}$ of the spot. We now give the origin of each term.

\subsubsection{The solid path \texorpdfstring{$g_{\mathrm{spot}}$}{g\_spot}}\label{sec:2.3.1}

The face across which the solids meet is the circle in which the surfaces
of the two spheres intersect. With the centre-to-centre distance
$D=r_i+r_j-\delta$, the radius $a$ of the intersection circle is given by
\begin{equation}
  x_i=\frac{D^2+r_i^2-r_j^2}{2D},\qquad a^2=r_i^2-x_i^2 .
  \label{eq:contactradius}
\end{equation}
The resistance across this face is the series combination of the
constriction resistances $R_{\mathrm{con},i}=1/(4\kappa_ia)$ and
$R_{\mathrm{con},j}=1/(4\kappa_ja)$, which arise because the streamlines
inside each particle converge towards the face, and the interfacial
resistance of the face itself,
$R_{\mathrm{int}}=(R_{c,i}+R_{c,j})/(\pi a^2)$ (\cref{sec:2.1}):
\begin{equation}
  g_{\mathrm{spot}}=\Big[\frac{1}{4\kappa_ia}+\frac{1}{4\kappa_ja}
  +\frac{R_{c,i}+R_{c,j}}{\pi a^2}\Big]^{-1}.
  \label{eq:gspot}
\end{equation}
Throughout this paper, constriction resistance refers to the resistance
that arises as the streamlines inside a particle are drawn into the contact
spot. It is the thermal analogue of Holm's electrical constriction
resistance~\citep{holm1967}.

\subsubsection{The matrix annulus \texorpdfstring{$g_{\mathrm{ann}}$}{g\_ann}}\label{sec:2.3.2}

The matrix layer outside the intersection circle follows from the same
annular integral as the near-contact branch with $h$ replaced by $-\delta$
and the lower limit moved from $r=0$ to $r=a$ (the derivation is in
\cref{app:A.3}), giving
\begin{equation}
  g_{\mathrm{ann}}=2\pi \kmat\rt\,
  \ln\!\Big(1+\frac{\rt-\delta}{a_{K,i}+a_{K,j}+h_{\mathrm{BOB}}}\Big).
  \label{eq:gann}
\end{equation}
Because the lower limit $a$ does not appear in the closed form, the
conductance of the matrix annulus does not depend on the contact radius.

\subsection{The bond branch}\label{sec:2.4}

A pair of particles belonging to the same continuous solid has no
interface. Bond branches represent the internal edges of an aggregate;
particles that are merely close together in space remain contact or
near-contact pairs. The constriction resistance of the contact branch must
not be carried over to them. Carrying it over would make the resistance of
the whole chain grow in proportion to the number of subdivisions, so that
the discretisation step would fix a physical quantity (\cref{app:A.4}).
What remains is axial bulk conduction through the cross-section,
\begin{equation}
  g_{\mathrm{bond}}=\frac{\kappa_h\,\pi R_{\min}^2}{L},
  \qquad R_{\min}=\min(r_i,r_j),
  \label{eq:gbond}
\end{equation}
where $L$ is the centre-to-centre distance; in the network of
\cref{fig:3}(c) this edge is written
$R_{\mathrm{axial}}=1/g_{\mathrm{bond}}$. A pair belonging to the same
solid but separated ($h>0$) is treated as a near-contact branch rather than
a bond branch, since the solid is not continuous there.

\subsection{The through-particle resistance}\label{sec:2.5}

All three kinds above describe only the neighbourhood of the surfaces of a
pair, and none of them contains the resistance to heat passing through the
particle itself. To represent this resistance while keeping the vertices
isothermal, we add on each edge, in series,
\begin{equation}
  \Rthr=\frac{x_i}{\kappa_i\pi r_i^2}+\frac{x_j}{\kappa_j\pi r_j^2},
  \label{eq:rthr}
\end{equation}
where $x_i$ is the distance from the intersection circle defined in
\cref{sec:2.3} to the centre of the particle, and the cross-section used is
the great circle $\pi r_i^2$ of the particle rather than the intersection
circle, so as to allow the heat to diffuse three-dimensionally. On a
near-contact edge we set $x_i=r_i$. A bond edge is itself conduction inside
the solid, so this term is not added a second time. The term is in series
with the whole edge, that is, with the parallel combination of the spot and
the matrix annulus, and not inside the contact spot (\cref{app:A.4}).

\subsection{Conditions of validity}\label{sec:2.6}

The model is valid under two conditions. First, the contact and
near-contact branches exchange at $h=0$, so $g_e$ must not be discontinuous
there. In our model $x_i\to r_i$ holds exactly as $\delta\to0$
(\cref{app:A.1}); the path through the solid vanishes and the contact
branch converges to the same expression as the $h\to0$ limit of the
near-contact branch.

Second, the same region of matrix must not be counted by more than one
pair. We remove from the edge set every pair whose logarithmic factor
$\eta=\ln(1+\rt/h)$ on the near-contact branch falls below $\etacut=1$
(\cref{tab:1}), and use an absolute distance cut-off of $0.5d$ alongside it
as a safeguard (under the conditions of this paper the condition on
$\etacut$ always acts first). We call this truncation the interaction
cut-off. The condition only removes edges, however; it does not compensate
the conduction through the removed region by any other path. Since that
region grows as the packing fraction falls, the lower bound of the range of
applicability of the model is set here. The details are given in
\cref{sec:7.1}.

\subsection{The contribution of each element}\label{sec:2.7}

We confirm numerically how much each element of the rule contributes to the
result. The conditions we used are given in \cref{tab:1} and the results in
\cref{tab:2}. The thermal conductivities of the filler and the matrix are
reference values taken as representative of commercial grades. Of the
interfacial resistances, $R_s$ is the reciprocal of a measured interfacial
thermal conductance~\citep{yang2021}, and the surface treatment is
expressed as a difference in this value. For the interfacial thermal
resistance $R_c$ between particles no established literature value exists,
so we assume a value based on the authors' experience and evaluate in
\cref{sec:7.3} the degree to which its uncertainty propagates to the
result.

\begin{table}[tbp]
  \centering
  \caption{Computation conditions.}\label{tab:1}
  \small
  \begin{tabular}{@{}p{0.26\linewidth}p{0.66\linewidth}@{}}
    \toprule
    & Value \\
    \midrule
    Filler conductivity $\kappa$ & AlN 170, Al$_2$O$_3$ 30, ZnO 25,
      h-BN 30 (through-plane), diamond 2000, SiO$_2$ $1.4\WmK$ \\
    Matrix conductivity $\kmat$ & epoxy 0.20, silicone 0.25,
      high-conductivity matrix $1.00\WmK$ \\
    Untreated & $R_s=1.11\times10^{-7}$,
      $R_c=1.0\times10^{-8}\mKW$ \\
    Surface-treated & $R_s=3.80\times10^{-8}$,
      $R_c=2.0\times10^{-9}\mKW$ \\
    Idealised & $R_s=0$ (only for the comparison with effective medium
      theory in \cref{sec:5.2}; 1{,}026 of the 3{,}078 conditions) \\
    Particle diameter $d$ / packing fraction $\phi$
      & $1$--$100\um$ / $0.37$--$0.70$ (random packing; the bimodal case of
      \cref{sec:6} has actual diameters $3.17$--$79.32\um$) \\
    Interaction cut-off & $\etacut=1$, absolute distance cut-off $0.5d$
      ($\etacut$ always acts first) \\
    Contact radius / interphase & geometric intersection circle / none
      ($t_{\mathrm{ip}}=0$; the hypothetical case \texttt{lowk\_coat} of
      \cref{app:C} is the only exception) \\
    Boundary slab thickness & 0.06 of the specimen length \\
    Virtual terminal & $\gvirt=10^3\max_e g_e$ (used only for evaluating
      $\Geff$) \\
    \bottomrule
  \end{tabular}
\end{table}

\begin{table}[tbp]
  \centering
  \caption{The contribution of each element to the result.}\label{tab:2}
  \footnotesize
  \begin{tabular}{@{}p{0.15\linewidth}p{0.19\linewidth}p{0.25\linewidth}
                    p{0.10\linewidth}p{0.19\linewidth}@{}}
    \toprule
    Element & What was quantified & Result & Runs & Consequence \\
    \midrule
    Contact branch (\cref{sec:2.3})
      & Whether the matrix annulus or the solid path is the dominant path
      & In 89.4\% of the conditions the annulus carries more than 50\% of
        the edge & 2{,}646 & A representation with the spot alone misses
        the dominant path \\
    Through-particle resistance (\cref{sec:2.5})
      & The change in the effective conductance when it is added
      & A factor of $0.455$--$0.996$ (family medians $0.734$--$0.903$)
      & 180 & Typically just under 20\%; a halving for SiO$_2$ \\
    Interaction cut-off (\cref{sec:2.6})
      & The effect of relaxing the condition ($\etacut=0.1$)
      & The edge count rises from 0--18 to 54--176 and the effective
        thermal conductivity becomes 3.1--10.6 times effective medium
        theory~\citep{nan1997}
      & 3 (FCC lattice, $\phi=0.2$--$0.6$)
      & A premise for the validity of the model, not a range within which
        accuracy may be tuned \\
    \bottomrule
  \end{tabular}
\end{table}

We read three things from \cref{tab:2}. First, in 89.4\% of the conditions
the matrix annulus carries more than 50\% of the edge, so representing the
contact branch by the spot alone misses the path that carries most of the
heat. Second, the effect of adding the through-particle resistance is
typically just under 20\%, and a halving for SiO$_2$. Third, relaxing the
interaction cut-off from $\etacut=1$ to $0.1$ raises the edge count on the
same regular lattice from 0--18 to 54--176 and makes the effective thermal
conductivity 3.1--10.6 times effective medium theory ($\phi=0.2$--$0.6$),
because summing over distant pairs counts the same volume of matrix
repeatedly. The third row is a premise for the validity of the model, not a
range within which accuracy may be tuned.

\section{The conductance filtration and sink-relative persistence}\label{sec:3}

\subsection{Choosing the filtration function}\label{sec:3.1}

\Cref{sec:2} fixed the thermal conductance $g_e$ of each particle pair.
From here on we use this weight to describe the packed structure. We use
filtrations of weighted graphs and degree-zero persistent homology, which
are already applied in a variety of fields~\citep{aktas2019}. To give a
filtration is to give a pair consisting of a complex and a function, and
changing the function gives a different filtration. The candidates for the
function are a geometric quantity determined by the arrangement of the
particles alone, or the thermal conductance $g_e$ fixed in \cref{sec:2}. In
this section we first decide which of the two to take
(\crefrange{sec:3.1}{sec:3.2}), then examine the conditions under which the
two axes give the same ordering (\cref{sec:3.3}), and finally fix what is
counted on that axis (\crefrange{sec:3.4}{sec:3.6}).

\begin{figure}[tbp]
  \centering
  \includegraphics[width=\linewidth]{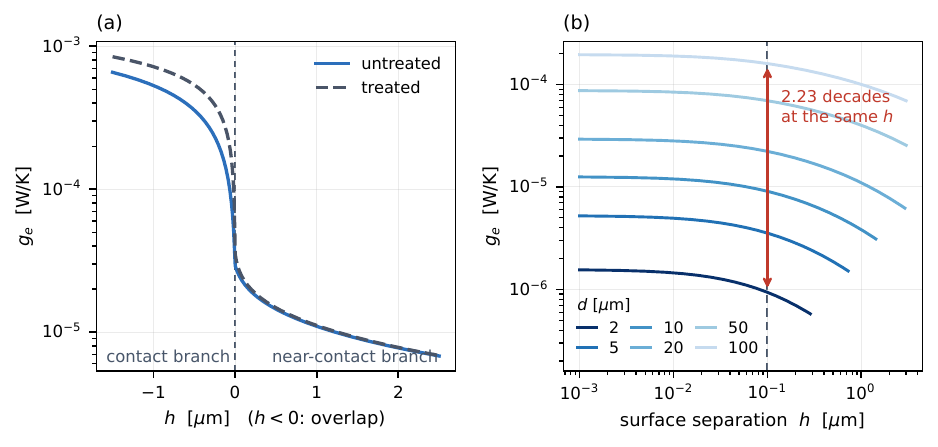}
  \caption{(a) $g_e$ against the surface separation $h$ for a pair of equal
  spheres of diameter $20\um$: the contact and near-contact branches join
  continuously at $h=0$. (b) The same curve for diameters from $2$ to
  $100\um$: at $h=0.1\um$ the curves span 2.23 decades, so $g_e$ is not a
  function of $h$ alone.}\label{fig:4}
\end{figure}

The choice of function matters because $g_e$ and a geometric quantity do
not induce the same ordering. For the pair of equal spheres of diameter
$20\um$ in \cref{fig:4}(a), the contact branch for $h<0$ and the
near-contact branch for $h>0$ join continuously at $h=0$, and $g_e$ is a
monotonically decreasing function of $h$. If instead the material is held
fixed and the particle diameter is varied from $2$ to $100\um$ as in
\cref{fig:4}(b), the six curves span 2.23 decades at one and the same
$h=0.1\um$. Hence $g_e$ is not a function of $h$ alone but of
$(h,\rt,\text{material})$, and the ordering of edges by $h$ and the
ordering by $g_e$ do not agree in general.

\begin{figure}[htbp]
  \centering
  \includegraphics[width=0.72\linewidth]{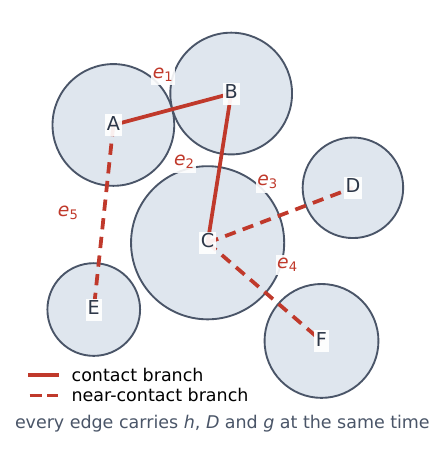}
  \caption{One edge carries three weights at once. The complex is fixed by
  the interaction cut-off; only the function on it is
  changed.}\label{fig:5}
\end{figure}

In what follows we use the centre-to-centre distance $D$ as the geometric
axis. The other geometric quantity, the surface separation, does not give
an ordering at the packing fractions we consider, because most edges have
$h\le0$, and we do not use it as an axis. The comparison is made on one and
the same edge set. The edge set is fixed geometrically by the interaction
cut-off of \cref{sec:2.6}; we hold the complex fixed and change only the
function. The way in which a single edge carries values of $h$, $D$ and
$g_e$ at the same time is shown in \cref{fig:5}.

\subsection{Introducing the conductance filtration}\label{sec:3.2}

This paper adopts the thermal conductance $g_e$ of a particle pair as the
filtration function, and calls it the conductance filtration. We set
\begin{equation}
  X_g=\{e\in E:\ g_e\ge g\}
  \label{eq:superlevel}
\end{equation}
and track the change of the connected components as the threshold $g$ is
decreased. In what follows we measure this threshold on the logarithmic
axis
\begin{equation}
  \theta=-\log_{10}g,
  \label{eq:theta}
\end{equation}
so that an edge with larger $g$ has smaller $\theta$. Since $g$ has the
dimension of $\mathrm{W/K}$, the absolute value of $\theta$ depends on the
system of units, and only differences in $\theta$ are unit-independent. We
therefore use the word ``decade'' for differences only, and whenever an
absolute value is stated we give the corresponding value of $g$ alongside
it. What this paper introduces is this function; it is neither the
superlevel set filtration as a construction nor $H_0$ as an invariant.
Following previous work, we call a quantity extracted from the barcode and
used to compare structures a descriptor. Taking $g_e$ as the axis makes the
descriptor a function of more than the structure alone: for one and the
same arrangement, changing the filler species or the surface treatment
changes the barcode.

\subsection{When the two axes are order isomorphic}\label{sec:3.3}

We fix two quantities for the comparison. The first is the ordering of the
edges, quantified by Kendall's rank correlation $\tau(D,\theta)$ between
the rank by the centre-to-centre distance $D$ and the rank by $\theta$
($\tau=1$ means the two orderings agree completely and $\tau=-1$ that they
are exactly reversed). The second is one scalar extracted from each axis.
We add edges in order of decreasing weight on the axis and take the value
of the weight at which the hot-side and the cold-side boundaries first
become connected. On the conductance axis we write this value $\thcrit$ and
call it the critical conductance level; on the centre-to-centre-distance
axis we write the same value $D^{*}$. Material values and analysis
parameters for the numerical examples follow \cref{tab:1}, and the
population is stated in the text and in the figure captions.

With these two we examine in turn the conditions under which the two axes
give the same ordering and those under which it breaks down.

\begin{figure}[tbp]
  \centering
  \includegraphics[width=\linewidth]{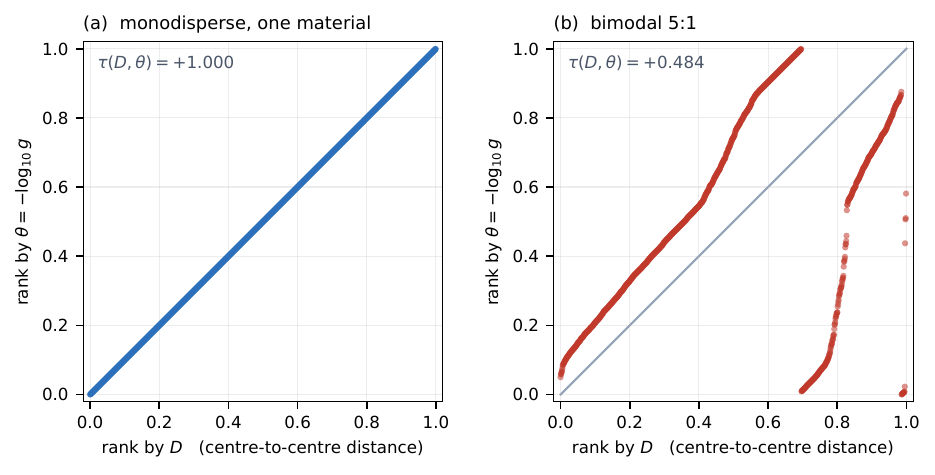}
  \caption{Rank by the centre-to-centre distance $D$ against rank by
  $\theta$. The material is the same in both panels and only the size
  distribution differs. (a) Monodisperse: every edge lies on the diagonal,
  so the two orders coincide. (b) With a particle size distribution the
  points leave the diagonal and separate into branches. Kendall's
  $\tau(D,\theta)$ is given in each panel.}\label{fig:6}
\end{figure}

First, there are cases in which the two axes agree completely. If all edges
have the same reduced radius $\rt$ and the material is uniform, then $g_e$
is a monotonically decreasing function of $h$ (\cref{sec:3.1}), and for
equal spheres $D=2r+h$, so the ordering by $D$ and the ordering by $\theta$
agree completely. In \cref{fig:6}(a) all edges of one structure lie on the
diagonal.

Second, agreement fails once the particle sizes are distributed. $g_e$
depends also on $\rt=r_ir_j/(r_i+r_j)$, so two edges with the same surface
separation have different $g_e$ if one joins large-diameter particles and
the other small-diameter ones. Changing only the particle size distribution
within the same group of structures gives $\tau=0.461$ for a bimodal 5:1
distribution and $\tau=-0.216$ for a broad continuous distribution,
reversing even the sign, and the points separate into two branches
corresponding to the two size classes (\cref{fig:6}(b)).

\begin{figure}[tbp]
  \centering
  \includegraphics[width=0.62\linewidth]{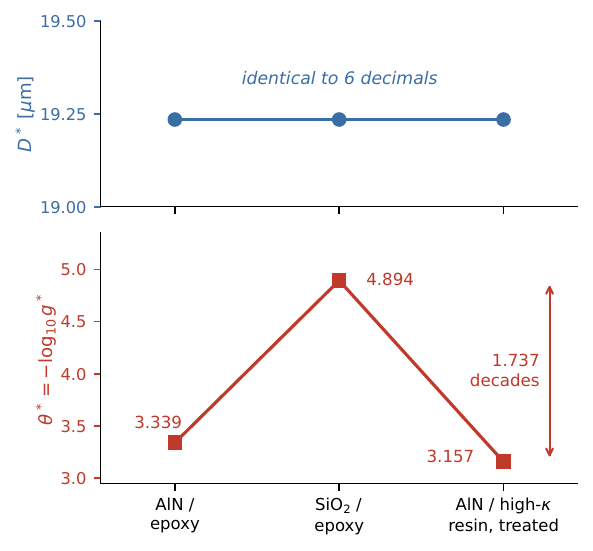}
  \caption{The same arrangement with only the filler species and the
  surface treatment changed. The geometric descriptor is unchanged while
  the critical conductance level spans 1.737 decades.}\label{fig:7}
\end{figure}

Third, when the material is involved, a difference that cannot appear on a
geometric axis in principle arises in the value of the descriptor itself.
For three conditions differing only in the filler species and the surface
treatment on one and the same arrangement, $D^{*}$ agrees to six decimal
places ($19.235295\um$), whereas $\thcrit$ takes $3.339$, $4.894$ and
$3.157$ -- the corresponding $\gstar$ being $4.6\times10^{-4}$,
$1.3\times10^{-5}$ and $7.0\times10^{-4}\WK$ -- and thus spans 1.737
decades (\cref{fig:7}).

These three are not exclusive. No a priori criterion gives the extent to
which each effect acts, so it is necessary to compute $g_e$ and compare the
ordering of the edges.

\subsection{Persistence relative to the sink boundary}\label{sec:3.4}

\begin{figure}[tbp]
  \centering
  \includegraphics[width=\linewidth]{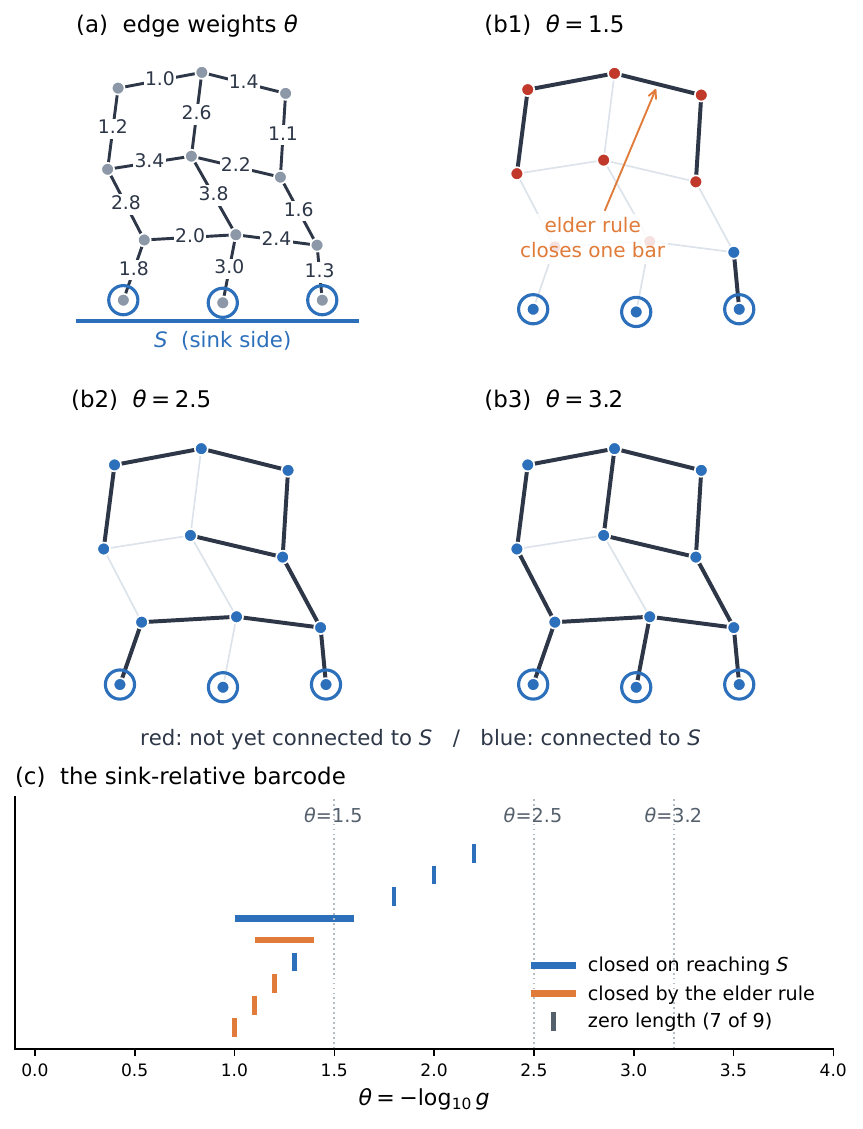}
  \caption{The construction of persistence relative to the sink boundary.
  A component dies at the moment it connects to $S$, so the length of a bar
  is the interval of $\theta$ over which that component was disconnected
  from the sink boundary.}\label{fig:8}
\end{figure}

Even once the function is fixed, what is counted on that function has to be
fixed separately. The absolute construction on $H_0(X_\theta)$ counts the
connectivity of the particles among themselves, whereas what carries
meaning in a heat transport problem is whether they have reached the
cold-side boundary, which we call the sink boundary below. The stages of
the construction are shown in \cref{fig:8}.

Let $S$ be the set of particles whose centres lie in the boundary slab on
the sink side. Using the axis $\theta=-\log_{10}g$ of \cref{sec:3.2}, we
define, for a threshold $\theta$,
\begin{equation}
  X_\theta=\{e\in E:\ -\log_{10}g_e\le\theta\}\cup S,
  \qquad M_\theta=H_0(X_\theta,\,S),
  \label{eq:relative}
\end{equation}
that is, we add edges in order of decreasing strength. $M_\theta$ is the
degree-zero homology of the space in which $S$ is contracted to a point,
and it counts only those connected components that have not reached $S$. As
$\theta$ increases, edges are added and at some time the component connects
to $S$ and dies. Collecting these birth--death pairs gives persistence
relative to the sink boundary. The length of a bar represents the interval
of $\theta$ over which that component was disconnected from the sink
boundary.

The hot-side source boundary is not included in $S$. Were it included, the
source and sink boundaries would be identified and the time at which the
two first become connected would no longer appear as a death event, so the
information about percolation would be lost. Taking $S$ to be the sink side
only, the death time of the component containing the source boundary gives
the critical conductance level of \cref{sec:3.6}.

In the implementation, rather than contracting $S$ to a point, we add a
virtual terminal vertex on the source side and another on the sink side and
attach edges of sufficiently large conductance $\gvirt$ (the value is in
\cref{tab:1}) between each terminal and the particles of the corresponding
boundary slab. Exact identification is the limit $\gvirt\to\infty$; at
finite $\gvirt$ a bar arises whose birth time is that value. This
originates in the method of implementation and is not a property of the
structure.

\subsection{Equivalent representations and stability}\label{sec:3.5}

Our construction gives the same barcode as the absolute persistent homology
of the quotient graph in which $S$ is contracted to a point, and the time
at which the component of a vertex $v$ reaches $S$ equals the bottleneck
value of the widest path from $v$ to $S$ (\cref{app:B.1}). Mergers between
components not containing $S$ are matched by the elder rule. Under this
convention the construction is a persistence module on a fixed
complex~\citep{chazal2016}; the dimension at each $\theta$ is finite, so an
interval decomposition exists and the barcode is uniquely determined.

Applying the stability theorem~\citep{cohensteiner2007} requires that the
subspace $S$ used for the relativisation and the complex, that is the edge
set, both be fixed. The former holds by construction. The latter depends on
the kind of perturbation. A perturbation that changes only the edge weights
$g_e$ preserves the complex, so the theorem applies, and when the
logarithms of the edge weights are changed from $\log g$ to $\log g'$ the
bottleneck distance $\dB$ between the barcodes satisfies
\begin{equation}
  \dB\le\|\log g-\log g'\|_\infty .
  \label{eq:stability}
\end{equation}
A perturbation that moves the particle coordinates, on the other hand,
causes edges to enter and leave because the interaction cut-off of
\cref{sec:2.6} determines the edge set from the coordinates, so the complex
changes. That case falls outside the premises of the theorem and carries no
guarantee. How useful the bound is as an estimate is quantified in
\cref{sec:7.2}.

\subsection{The critical conductance level and the descriptor}\label{sec:3.6}

What we use from the barcode as a descriptor is the $\theta$ at which the
source and sink boundaries first become connected, that is the critical
conductance level $\thcrit$ fixed in \cref{sec:3.3}, and we write $\gstar$
for the corresponding value of the conductance. By the construction of
\cref{sec:3.4}, the death time of the component containing the source
boundary coincides exactly with $\thcrit$. $\gstar$ is the height of the
bottleneck of the path that percolates the network, and it gives the
minimum level of conductance the structure in question requires in order to
transport heat. The length of the longest bar differs from $\thcrit$ only
by a constant, and the bar originating in the virtual terminal is an
artefact of the implementation, so neither is counted as an independent
descriptor.

\begin{figure}[tbp]
  \centering
  \includegraphics[width=0.86\linewidth]{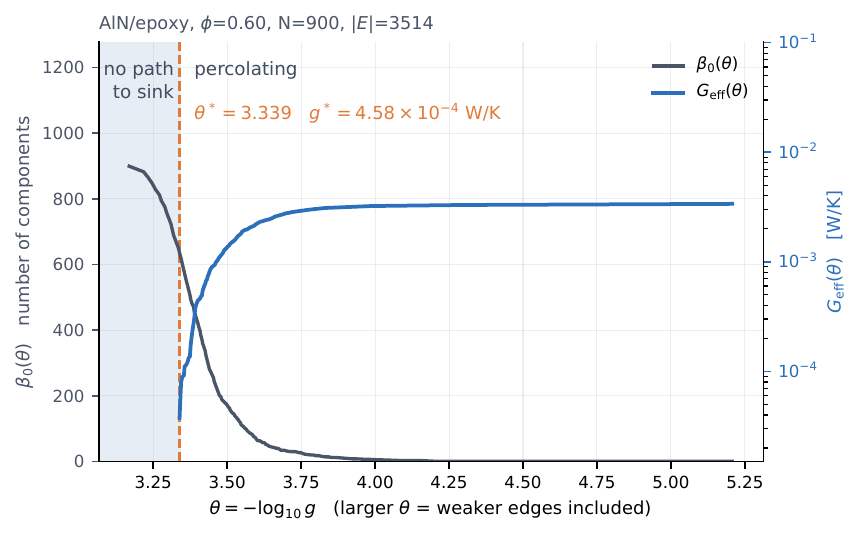}
  \caption{A sweep in $\theta$ for a single structure of equal spheres
  ($d=20\um$, $\phi=0.600$, AlN / epoxy / untreated, $N=900$). The first
  percolating path appears at $\thcrit=3.339$
  ($\gstar=4.58\times10^{-4}\WK$), while more than 600 components
  remain.}\label{fig:9}
\end{figure}

A numerical example is shown in \cref{fig:9}. For a single structure of
equal spheres ($d=20\um$, $\phi=0.600$, AlN / epoxy / untreated, $N=900$)
we obtain $\thcrit=3.339$ ($\gstar=4.58\times10^{-4}\WK$), and below this
$\theta$ no percolating path exists. Percolation does not imply that the
whole is connected: past $\thcrit$ the number of components $\beta_0$
remains above 600 and keeps falling until $\theta\approx4.2$. The $\theta$
at which the first path appears and the $\theta$ at which the whole becomes
connected are different quantities, and what this paper treats as $\thcrit$
is the former.

$\gstar$ is the same quantity as the critical conductance of critical path
analysis in transport through disordered systems~\citep{ahl1971}. That
paper writes the conductivity as a ratio of the critical conductance to a
characteristic length of the network and leaves the characteristic length
as an unknown quantity; for the same reason, our estimate of $\Geff$ from
$\thcrit$ remains accurate only to about a decade (\cref{sec:5.3}).

The critical conductance level $\thcrit$ and the corresponding $\gstar$ do
not, however, return which edges contribute how much to the effective
thermal conductivity. That is given by the dissipation share of
\cref{sec:4}.

\section{The dissipation share as a logarithmic sensitivity}\label{sec:4}

\subsection{Why the temperature field must be solved}\label{sec:4.1}

\Cref{sec:3} described the structure using the edge weights $g_e$ alone.
What that yields is the level of the bottleneck to percolation
(\cref{sec:3.6}), not which edges are limiting the effective thermal
conductivity.

This limitation is essential. $g_e$ can be computed from the geometry and
the material properties without solving the network, whereas the
temperature difference $(\mathrm du)_e$ between the two ends of an edge is
obtained only by solving the temperature field under the boundary
conditions. No heat flows along a weak edge, so the magnitude of $g_e$ does
not imply the magnitude of the transport the edge carries. Below we show
what can be claimed once the temperature field is solved.

\subsection{Setting}\label{sec:4.2}

We define the effective conductance $\Geff$ as the minimum of the Dirichlet
energy,
\begin{equation}
  \Geff(g)=\min_{u:\ u|_{\mathrm{src}}=1,\ u|_{\mathrm{snk}}=0}\
  \sum_{e\in E}g_e\,(\mathrm du)_e^2 .
  \label{eq:dirichlet}
\end{equation}
Here $u$ is the temperature field on the vertices and $(\mathrm du)_e$ the
difference between the two ends of edge $e$. The right-hand side is a
convex quadratic form in $u$, and its stationarity condition is that the
balance of heat flow vanish at every vertex whose temperature is not fixed
-- the current law -- so the minimiser $u^\star$ is nothing but the steady
temperature field under the boundary conditions of \cref{sec:2.1}. With the
boundary temperature difference normalised to unity, this minimum takes the
same value as the total heat flow from source to sink in the steady field,
namely the $\Geff$ of \cref{sec:2.1}.

Writing the minimiser as $u^\star$, we define the dissipation of edge $e$ as
$P_e=g_e(\mathrm du^\star)_e^2$. Identifying temperature with potential,
this is the Joule dissipation consumed by the edge in an electrical
circuit. We define its fraction
\begin{equation}
  \Pshare_e=\frac{P_e}{\Geff}
  \label{eq:pshare}
\end{equation}
and call it the dissipation share. Because the boundary conditions are
normalised to $1$ and $0$, $\Geff$ equals the total dissipation. Since the
implementation uses virtual terminals of finite conductance
(\cref{sec:3.4}), the dissipation share is normalised over the physical
edges only in the numerical computation.

\subsection{The dissipation share equals the logarithmic sensitivity}\label{sec:4.3}

Four things can be said about the dissipation share. The main result of
this paper is the third of them.

\textbf{Proposition 1 (envelope theorem).} The partial derivative of the
effective conductance with respect to each edge weight equals the square of
the temperature difference at the minimiser (\cref{app:B.2}):
\begin{equation}
  \frac{\partial \Geff}{\partial g_e}=(\mathrm du^\star)_e^2 .
  \label{eq:envelope}
\end{equation}
Although the minimiser $u^\star$ itself depends on $g$, its contribution
does not appear in the derivative. This is because $u^\star$ is an interior
stationary point, that is, because the current law holds at every vertex
that is not fixed.

\textbf{Proposition 2 (sum rule).} The sum of the dissipation equals the
effective conductance (\cref{app:B.2}):
\begin{equation}
  \sum_{e\in E}P_e=\Geff .
  \label{eq:sumrule}
\end{equation}
This is an identity that follows from the definitions of \cref{sec:4.2}.

\textbf{Proposition 3 (logarithmic sensitivity).} Combining Propositions 1
and 2 gives (\cref{app:B.2})
\begin{equation}
  \Pshare_e=\frac{\partial\log \Geff}{\partial\log g_e},
  \label{eq:logsens}
\end{equation}
that is, the dissipation share equals exactly the relative response of the
effective conductance to a relative change of the conductance of that edge.

\textbf{Proposition 4 (concavity).} $\Geff$ is monotonically increasing and
concave in each $g_e$ (\cref{app:B.3}). Improving the same edge repeatedly
therefore has diminishing returns.

We quantified how far the ordering of the edges by $g_e$ actually differs
from the ordering by the dissipation share, over 69 packings formed from 23
structure families -- varying the packing fraction, the particle size
distribution, defects, process-induced segregation and fibre anisotropy --
with three random seeds each (the generation rules and the generated
properties of the families are in \cref{app:C}). The selection efficiency
$\varepsilon$ is the gain obtained by selecting the top 20 edges in order
of $g_e$ and improving them, divided by the gain obtained by selecting in
decreasing order of the dissipation share. The results are given in
\cref{tab:3}.

\begin{table}[tbp]
  \centering
  \caption{The ordering by $g_e$ compared with the ordering by the
  dissipation share, over 69 packings (23 families, three seeds
  each).}\label{tab:3}
  \begin{tabular}{@{}p{0.62\linewidth}r@{}}
    \toprule
    Quantity compared & Result \\
    \midrule
    Spearman correlation between the two ranks (median over packings)
      & 0.315 \\
    Number of the top 20 edges shared (median)
      & 1 (0 in 33 of the 69) \\
    Selection efficiency $\varepsilon$ of the ordering by $g_e$ (median)
      & 0.083 \\
    Packings with $\varepsilon<0.5$ & 61/69 \\
    Win--loss record against a random selection & 35 to 34 \\
    \bottomrule
  \end{tabular}
\end{table}

The degree of overlap was measured by four statistics: the Spearman
correlation between the two ranks, the number of the top 20 edges shared,
the selection efficiency $\varepsilon$, and the win--loss record against a
random selection (the comparison against a random selection is made against
the mean of 20 random selections per packing). All four show that the
ordering by $g_e$ scarcely overlaps the ordering by the dissipation share,
and that a selection based on $g_e$ is indistinguishable from a random
selection. What orders the edges is therefore not $g_e$ but $\Pshare$, and
Proposition 3 states that this ordering is an exact ordering by
sensitivity.

\subsection{An exact closed form for a finite change}\label{sec:4.4}

The preceding subsection concerns a derivative, that is, an infinitesimal
change. A closed form is available for a finite change as well.

When the conductance of edge $e=(i,j)$ is changed as $g_e\to g_e+\Delta$,
the identity
\begin{equation}
  \Geff(g+\Delta\mathbf e_e)=\Geff(g)
  +\frac{\Delta\,(\mathrm du^\star)_e^2}{1+\Delta\,\Rt_e}
  \label{eq:finitechange}
\end{equation}
holds exactly for any $\Delta$ (\cref{app:B.3}). Here $\Rt_e$ is the
effective resistance between $i$ and $j$ in the circuit in which the source
and the sink are shorted to a single point. This is an exact equality and
not a first-order approximation; it tends to the derivative of
\cref{sec:4.3} as $\Delta\to0$ and to the value for a shorted edge as
$\Delta\to\infty$. In a response under boundary conditions the perturbation
acts on the quotient space in which source and sink are identified, so
substituting the ordinary effective resistance of the circuit does not
satisfy the equality. The derivation is a rank-one perturbation of the
Laplacian (\cref{app:B.3}).

Since $\Rt_e$ is obtained with one back-substitution per edge once the
shorted Laplacian has been factorised, evaluating a finite change only for
the top-ranked edges by $\Pshare$ is feasible at a realistic computational
cost without re-solving.

\subsection{The idealisation ratio of each mechanism}\label{sec:4.5}

The discussion so far concerns a change of a single edge. A change by
mechanism -- lowering the interfacial resistance, or raising the thermal
conductivity of the matrix -- is what arises in design, and it changes many
edges at once, so the closed form of \cref{sec:4.4} does not apply. For a
mechanism $m$ we therefore define the idealisation ratio as the ratio to
the effective conductance under idealisation,
\begin{equation}
  \Gamma_m=\frac{\Geff\big(g^{(m\to\mathrm{ideal})}\big)}{\Geff(g)} .
  \label{eq:gamma}
\end{equation}
The idealisation is applied to each of three mechanisms separately: the
interface, setting both the contact-interface and the matrix-interface
thermal resistances to zero; the matrix, setting the thermal conductivity
of the matrix to infinity; and the particle, setting the thermal
conductivity of the particle to infinity.

$\Gamma_m$ is an upper bound under idealisation of the mechanism. A
mechanism whose $\Gamma_m$ is close to unity leaves the effective
conductance unchanged no matter how far it is improved, whereas a large
$\Gamma_m$ only indicates that room for improvement may exist, and the
ordering is reversed by the configuration (\cref{sec:6.2}).

\subsection{Relation to previous work}\label{sec:4.6}

Among indicators that assign a sensitivity-like quantity to the edges of a
graph there is the weight $g_eR_{\mathrm{eff}}(e)$ used in graph
sparsification~\citep{spielman2011}. That quantity is the logarithmic
derivative of the weighted number of spanning trees, which measures a
different object from the logarithmic derivative of the effective
conductance, so the two give different edges at the top in general.
Moreover, whereas previous persistent homology in materials science has
related descriptors to properties as a statistical correlation, $\Pshare$
is an exact expression for the response to a change of the conductance.

\section{Implementation, computational cost, and verification}\label{sec:5}

In this section we cover the implementation of the procedures of
\cref{sec:2,sec:3,sec:4} (\cref{sec:5.1}), the computational cost
(\cref{sec:5.2}), and the comparison with effective medium theory together
with the estimate of the effective conductance (\cref{sec:5.3}).

\subsection{Implementation}\label{sec:5.1}

In this subsection we follow the implementation of the procedures fixed in
\cref{sec:2,sec:3,sec:4} from the input through to the four procedures that
compute the quantities.

The input is the coordinates and radii of the particles together with the
assignment of materials. The processing consists of the neighbour search
and the classification of the edges -- enumerating candidate pairs with a
$k$-d tree, sorting them into contact, near-contact and bond by the two
conditions of \cref{sec:2.1}, and assigning $g_e$ by the rules of
\cref{sec:2} -- the construction of the graph with virtual terminals added
to the sink and source boundaries, and four procedures: (a) construction of
the barcode, (b) the critical conductance level, (c) the dissipation share
and (d) a finite change.

\begin{figure}[tbp]
  \centering
  \includegraphics[width=\linewidth]{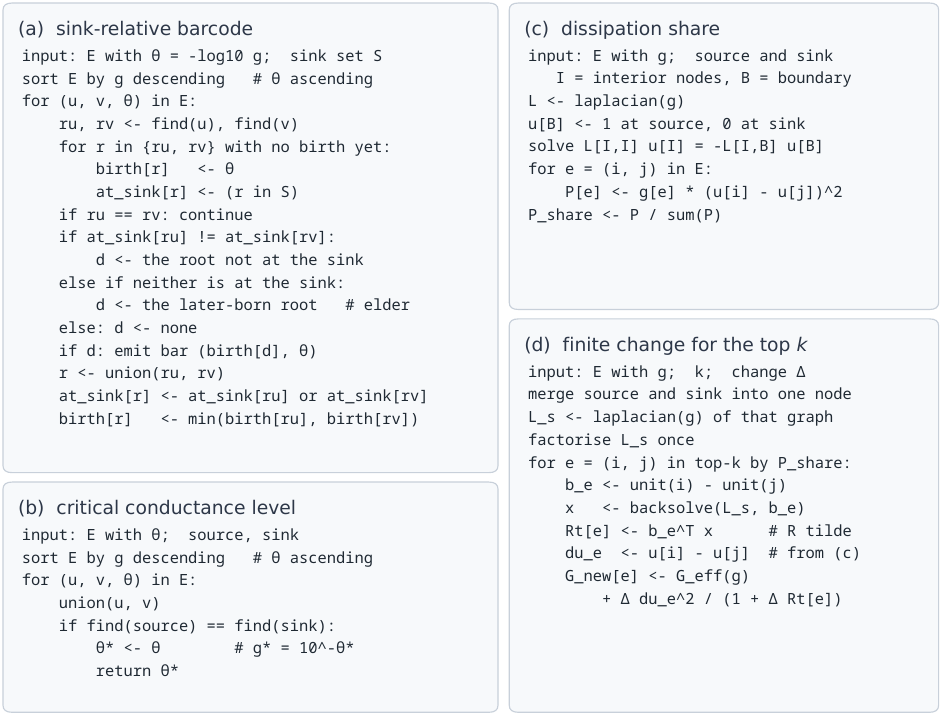}
  \caption{Pseudocode for the four procedures. Barcode construction is a
  single pass over the sorted edges; the critical conductance level is a
  second pass over the same ordering; the dissipation share is one sparse
  solve; a finite change is one factorisation followed by $k$
  back-substitutions.}\label{fig:10}
\end{figure}

All four are combinations of existing algorithms. Pseudocode is given in
\cref{fig:10} and the output and the cost of each procedure in
\cref{tab:4}.

Procedure (a) sorts the edges in decreasing order of conductance and
processes them in a single pass, merging components with union--find.
Procedure (b) is a pass over the same ordering that returns the $g$ at the
moment the source and the sink first lie in one component. Procedure (c)
solves the linear system on the interior vertices with the boundary
temperatures fixed by one sparse solve, obtains the dissipation
$g_e(\mathrm du)_e^2$ of all edges at once, and normalises by the sum.
Procedure (d) factorises once the shorted Laplacian in which source and
sink are identified, obtains $\Rt_e$ -- the effective resistance on the
shorted circuit defined in \cref{sec:4.4} -- for each of the top $k$ edges
with one back-substitution, and substitutes it into the closed form.

The implementation is released as a separate software record that
reproduces every figure of this paper; its identifier is given in the
ancillary material. Its settings, its verification checks and the contents
of the bundle are described there as well (the checks are items V1--V20 of
the same table).

\begin{table}[tbp]
  \centering
  \caption{The procedures and their computational cost
  ($\lvert E\rvert$ is the number of edges).}\label{tab:4}
  \small
  \begin{tabular}{@{}p{0.32\linewidth}p{0.26\linewidth}p{0.29\linewidth}@{}}
    \toprule
    Procedure & Output & Cost \\
    \midrule
    Preprocessing (neighbour search and edge classification)
      & $g_e$ for all edges
      & Dominated by the $k$-d tree, $O(N\log N)$ \\
    (a) Construction of the barcode & The bars of the relative persistence
      & Dominated by the sorting,
        $O(\lvert E\rvert\log\lvert E\rvert)$ \\
    (b) Critical conductance level & $\thcrit$ & One pass \\
    (c) Dissipation share & $\Pshare$ for all edges
      & One sparse solve and $O(\lvert E\rvert)$ \\
    (d) Finite change & $\Geff$ with the top $k$ edges changed
      & One factorisation and $k$ back-substitutions \\
    \bottomrule
  \end{tabular}
\end{table}

\subsection{Computational cost}\label{sec:5.2}

In this subsection we measure how the computation time grows with the
number of particles.

\begin{figure}[tbp]
  \centering
  \includegraphics[width=\linewidth]{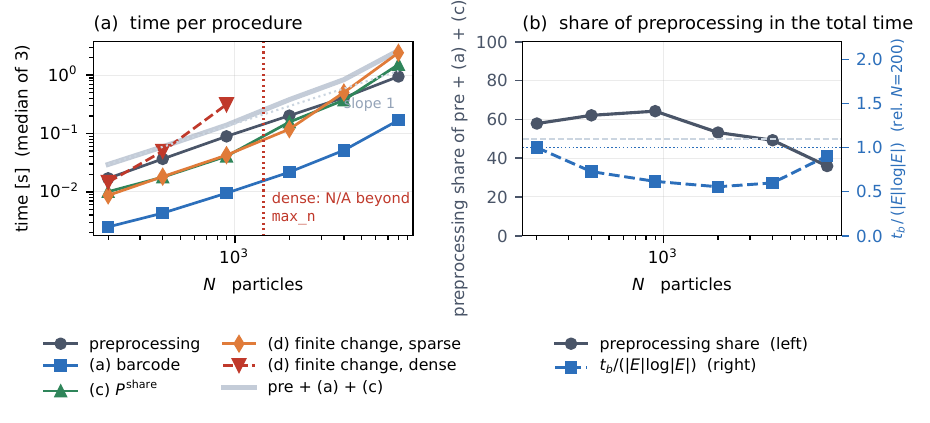}
  \caption{Timing against the number of particles, three runs at each $N$.
  (a) Time per procedure. (b) Left axis: the share of the preprocessing,
  which is the largest of the three stages up to $N\approx4000$, beyond
  which the single sparse solve of (c) overtakes it. Right axis: the
  barcode time divided by $|E|\log|E|$, normalised at $N=200$; the ratio
  stays within a factor $1.8$, so the sorting dominates.}\label{fig:11}
\end{figure}

We timed structures with the number of particles $N$ varying from 200 to
8{,}000, three times at each $N$. Material values and analysis parameters
follow \cref{tab:1}. The results are shown in \cref{fig:11}. The slopes on
the log--log plot are 1.08 for the preprocessing, 1.12 for the construction
of the barcode (a), 1.36 for the dissipation share (c) and 1.49 for the
evaluation of a finite change (d); all are close to linear in $N$. The
timings were measured on a single machine.

Of the time taken by the preprocessing and by procedures (a) and (c), the
preprocessing -- that is, the neighbour search and the classification of the
edges -- accounts for between 35.9\% and 64.2\%, and it is the largest of
the three up to $N\approx4000$; beyond that the single sparse solve of (c)
overtakes it. Neither stage is superlinear. Procedure (d) is left out of
this fraction because it is evaluated for a fixed number of edges ($k=20$)
rather than for all of them. The construction time of (a) divided by
$\lvert E\rvert\log\lvert E\rvert$ is nearly constant, consistent with the
cost given in \cref{tab:4}.

\subsection{Comparison with effective medium theory and an estimate of the effective conductance}\label{sec:5.3}

In this subsection we carry out two things: a comparison with effective
medium theory as a mean-field quantity, and an estimate of the level of the
effective conductance from $\thcrit$.

Effective medium theory~\citep{nan1997} is a mean-field theory for a
dispersed system that contains no contact between particles, and the
comparison is measured by the relative difference in $k_{\mathrm{eff}}$
(with effective medium theory in the denominator, in absolute value).
Material values follow \cref{tab:1}. The geometry is a homogenisation on
regular lattices with periodic boundaries -- neither the boundary slab nor
the interaction cut-off is used -- and $R_s=0$ (idealised) is added to the
levels. The conditions are given in \cref{tab:5}.

\begin{table}[tbp]
  \centering
  \caption{Conditions for the comparison with effective medium
  theory.}\label{tab:5}
  \small
  \begin{tabular}{@{}p{0.26\linewidth}p{0.66\linewidth}@{}}
    \toprule
    Item & Levels \\
    \midrule
    Lattice & simple cubic, SC ($z=6$, $\phi_{\max}=\pi/6$);
      body-centred cubic, BCC ($8$, $\pi\sqrt3/8$);
      face-centred cubic, FCC ($12$, $\pi/(3\sqrt2)$) \\
    Relative packing fraction $\phi/\phi_{\max}$
      & 0.67--1.00 (six or seven levels per lattice) \\
    Filler & SiO$_2$, ZnO, Al$_2$O$_3$, AlN, h-BN, diamond
      ($\kappa_f=1.4$--$2{,}000\WmK$) \\
    Matrix & silicone, epoxy, high-conductivity matrix
      ($\kmat=0.2$--$1.0\WmK$) \\
    Interfacial resistance $R_s$
      & 0 (idealised) / $3.8\times10^{-8}$ /
        $1.11\times10^{-7}\mKW$ \\
    Particle diameter $d$ & 1 / 10 / $100\um$ \\
    Total & 3{,}078 conditions (one third of them with $R_s=0$) \\
    \bottomrule
  \end{tabular}
\end{table}

\begin{figure}[tbp]
  \centering
  \includegraphics[width=\linewidth]{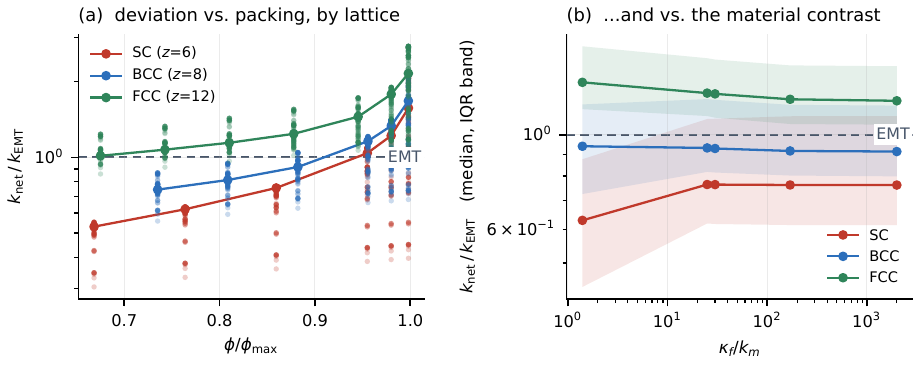}
  \caption{The deviation from effective medium theory. (a) The ratio
  $k_{\mathrm{net}}/k_{\mathrm{EMT}}$ by lattice against
  $\phi/\phi_{\max}$; the points are the individual conditions and the
  lines their medians. (b) The same ratio against the material contrast
  $\kappa_f/\kmat$, as medians and interquartile ranges.}\label{fig:12}
\end{figure}

We evaluated the ratio $k_{\mathrm{net}}/k_{\mathrm{EMT}}$ over all
3{,}078 conditions; \cref{fig:12}(a) plots it against $\phi/\phi_{\max}$ by
lattice and \cref{fig:12}(b) against the material contrast
$\kappa_f/\kmat$.

The median relative difference is 25.2\% and the median ratio is 1.07, with
no systematic bias. The median ratios by lattice are SC 0.76, BCC 0.93 and
FCC 1.23, which follow the order of the coordination number, and for every
lattice the ratio rises monotonically with $\phi/\phi_{\max}$, whereas it
is nearly flat for $\kappa_f/\kmat$ above 3 (only the 171 conditions at 1.4
give a ratio of 0.71). Effective medium theory is a mean field with no
picture of the paths, so this departure is the contribution of the path
structure that the coordination number and the packing fraction fix.

\begin{figure}[tbp]
  \centering
  \includegraphics[width=\linewidth]{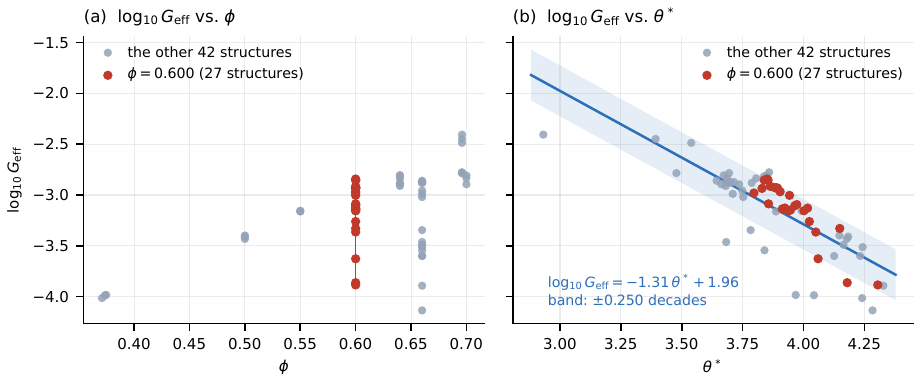}
  \caption{$\phi$ does not fix $\Geff$, whereas $\thcrit$ estimates its
  level. (a) $\log_{10}\Geff$ against $\phi$: the 27 packings at
  $\phi=0.600$ alone span 1.0 decade. (b) The same 69 packings against
  $\thcrit$.}\label{fig:13}
\end{figure}

For the estimate of $\thcrit$ we use the 69 packings of \cref{sec:4.3}
(\cref{app:C}). Among the 27 packings of spherical filler at $\phi=0.600$
alone, $\log_{10}\Geff$ spans 1.0 decade, so $\Geff$ is not fixed by $\phi$
alone (\cref{fig:13}(a)). Ordering the same 69 packings by $\thcrit$
instead, the points, spread over 1.7 decades, gather about the regression
line $\log_{10}\Geff=-1.31\,\thcrit+1.96$ (\cref{fig:13}(b)). The width of
the band, $\pm0.250$ decades, is the residual of the leave-one-family-out
cross-validation, and is smaller than the 0.348 decades of a regression on
$\phi$ alone. $\thcrit$ is a quantity that measures the bottleneck of the
percolating path directly (\cref{sec:3.6}), and it accounts for the
variation that $\phi$ does not fix.

The coefficients depend on the family, however, and one family alone -- the
group of structures obtained by adding jitter to a lattice, whose
generation mechanism differs -- is fitted poorly. We therefore use the
relation as an empirical rule for estimating a level, not for the values of
its coefficients. The remaining 0.25 decades correspond to the geometric
factor that the correspondence of \cref{sec:3.6} leaves as an unknown
quantity.

\section{A numerical demonstration}\label{sec:6}

\subsection{Computation conditions}\label{sec:6.1}

In this section we apply the framework of \cref{sec:2,sec:3,sec:4,sec:5} to
a single family of structures. Of the four design variables listed in
\cref{sec:1} we vary three -- the particle diameter, the filler species and
the surface treatment -- over eight conditions, holding the packing
fraction fixed at 0.700. There is no new theory here, and the structures
are synthetic rather than real formulations. The computation conditions are
given in \cref{tab:6}; material values and analysis parameters follow
\cref{tab:1}. We made the structure bimodal because a monodisperse packing
does not reach $\phi=0.700$ and because bimodal sizing is in practical use
as a means of raising the packing fraction. The upper nominal diameter is
$25\um$, because the large particles of a bimodal distribution are 3.17
times the nominal value and a larger nominal diameter would fall outside
the range of \cref{tab:1}.

\begin{table}[tbp]
  \centering
  \caption{Computation conditions for the demonstration.}\label{tab:6}
  \small
  \begin{tabular}{@{}p{0.32\linewidth}p{0.60\linewidth}@{}}
    \toprule
    & Value \\
    \midrule
    Structure & bimodal, size ratio 5:1 \\
    Number of particles $N$ / packing fraction $\phi$ & 2{,}000 / 0.700 \\
    Nominal diameter $d$ & $5\um$ (actual $3.17$--$15.86\um$),
      $25\um$ ($15.86$--$79.32\um$) \\
    Filler & AlN ($\kappa_f/\kmat=850$), SiO$_2$ ($\kappa_f/\kmat=7$) \\
    Matrix & epoxy ($\kmat=0.20\WmK$) \\
    Surface treatment & untreated / surface-treated \\
    Random seed & 3 \\
    Runs & 8 conditions $\times$ 3 seeds $=24$ \\
    Number of edges / contact number & 6{,}141--6{,}289 / 3.603--3.665
      (by seed) \\
    \bottomrule
  \end{tabular}
\end{table}

The principal subject of this paper is a filled polymer whose ratio of the
thermal conductivities of filler and matrix, $\kappa_f/\kmat$, lies between
$10^2$ and $10^4$~\citep{sarikhani2022}. AlN lies in this range, on the
side where the interfacial resistance is limiting, and SiO$_2$ lies below
the lower end and serves as a matrix-limited control. The four material
conditions sharing the same $d$ and the same seed are computed on one and
the same structure, with identical packing fraction, number of edges and
contact number, changing only the material.

The interventions applied to this group of structures, and the way they are
evaluated, are collected in \cref{tab:7}.

\begin{table}[tbp]
  \centering
  \caption{The interventions applied and the quantities
  evaluated.}\label{tab:7}
  \footnotesize
  \setlength{\tabcolsep}{4pt}
  \begin{tabular}{@{}p{0.16\linewidth}p{0.28\linewidth}p{0.23\linewidth}
                    p{0.22\linewidth}@{}}
    \toprule
    Item & Content & Quantity evaluated & Where \\
    \midrule
    Substitution of the material
      & Only the filler species is changed, on the same structure
      & $k_{\mathrm{eff}}$, contact share
      & \cref{tab:8}, \cref{fig:14}(a)--(c) \\
    Critical conductance level
      & $\thcrit$ for the eight conditions (neither structure nor material
        is changed; sorting and union--find only, with no Laplacian solved)
      & Shift with filler and with surface treatment
      & \cref{tab:8}, \cref{fig:14}(d) \\
    Application of the surface treatment
      & $R_s$ and $R_c$ are changed to the treated values, on the same
        structure
      & Ratio in $k_{\mathrm{eff}}$ (taken per seed, then the median)
      & \cref{tab:8}, \cref{fig:14}(a) \\
    Idealisation of a mechanism
      & Each mechanism is idealised in turn
      & Idealisation ratio $\Gamma_m$ (\cref{sec:4.5})
      & \cref{tab:8}, \cref{fig:14}(e) \\
    \bottomrule
  \end{tabular}
\end{table}

\subsection{Results}\label{sec:6.2}

\begin{figure}[tbp]
  \centering
  \includegraphics[width=\linewidth]{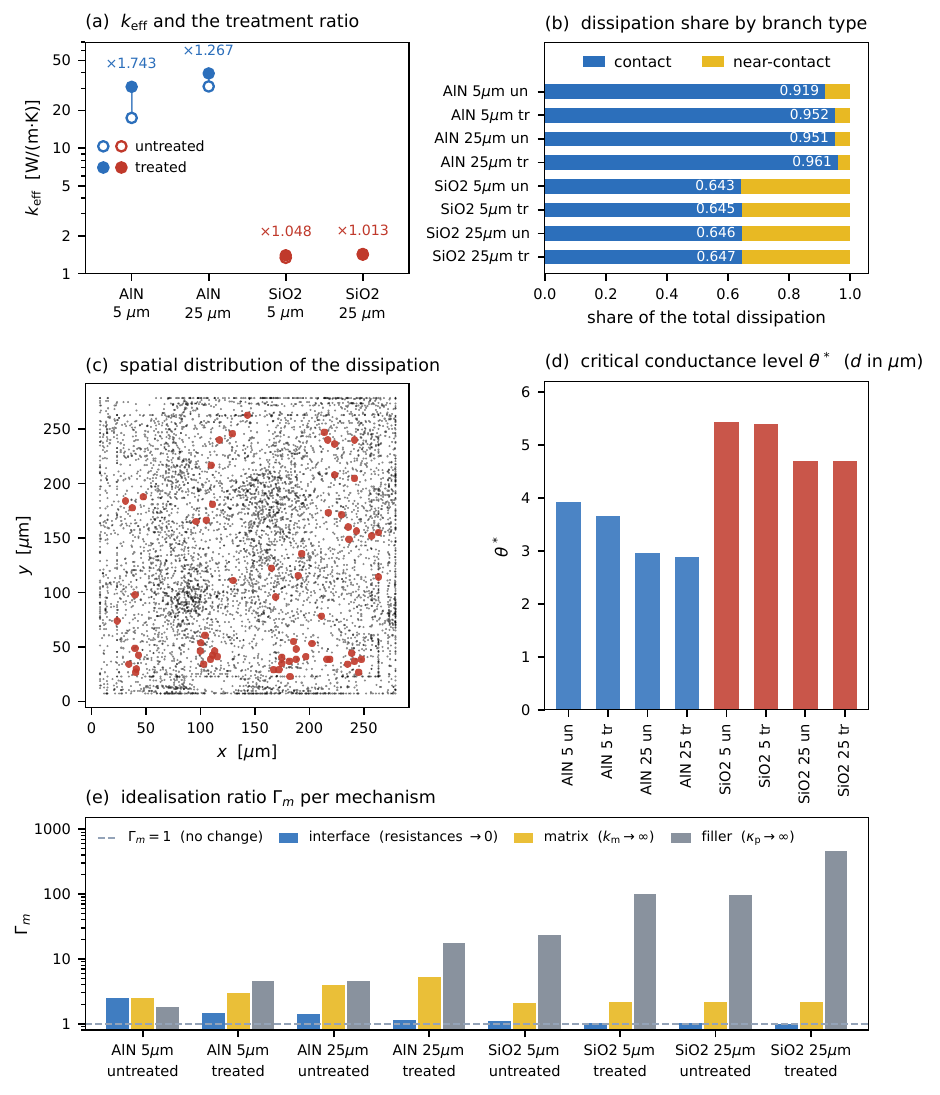}
  \caption{Numerical demonstration, in the order of the columns of
  \cref{tab:8}. (a) $k_{\mathrm{eff}}$ and the treatment ratio;
  (b) dissipation share by branch type; (c) spatial distribution of the
  dissipation for one run (AlN, untreated, $d=25\um$), with the top
  $1\%$ of the edges in red; (d) critical conductance level $\thcrit$; (e) idealisation
  ratio $\Gamma_m$ per mechanism.}\label{fig:14}
\end{figure}

The results of the interventions of \cref{tab:7} are given in \cref{tab:8}
and the figures in \cref{fig:14}. The values are medians over three
seeds. The contact share is the
fraction of the total dissipation carried by contact branches, and the
column headed ``half'' is the number of edges required to carry half of the
total dissipation, as a fraction of the number of edges. The treatment
ratio is taken per seed on one and the same structure and therefore does
not equal the quotient of the medians of $k_{\mathrm{eff}}$. The three
values of $\Gamma_m$ are in the order interface / matrix / particle.

\begin{table}[tbp]
  \centering
  \caption{Results for the eight conditions. Medians over three seeds.}
  \label{tab:8}
  \small\setlength{\tabcolsep}{4pt}
  \begin{tabular}{@{}r l r r r r r l@{}}
    \toprule
    $d$ [$\mu$m] & Filler, treatment & $k_{\mathrm{eff}}$
      & Ratio & Contact & Half [\%] & $\thcrit$ & $\Gamma_m$ \\
    & & [W/(m$\cdot$K)] & & share & & & \\
    \midrule
    5  & AlN, treated      & 30.81 & 1.743 & 0.952 & 0.83 & 3.665 & 1.44/2.97/4.62 \\
    5  & AlN, untreated    & 17.41 & ---   & 0.919 & 0.80 & 3.928 & 2.51/2.48/1.82 \\
    5  & SiO$_2$, treated  & 1.401 & 1.048 & 0.645 & 4.79 & 5.405 & 1.03/2.16/98.7 \\
    5  & SiO$_2$, untreated& 1.337 & ---   & 0.643 & 4.66 & 5.429 & 1.08/2.10/23.1 \\
    25 & AlN, treated      & 39.40 & 1.267 & 0.961 & 0.94 & 2.883 & 1.13/5.19/17.49 \\
    25 & AlN, untreated    & 31.03 & ---   & 0.951 & 0.83 & 2.965 & 1.44/3.91/4.60 \\
    25 & SiO$_2$, treated  & 1.435 & 1.013 & 0.647 & 4.87 & 4.693 & 1.01/2.19/466.3 \\
    25 & SiO$_2$, untreated& 1.417 & ---   & 0.646 & 4.84 & 4.700 & 1.02/2.18/97.8 \\
    \bottomrule
  \end{tabular}
\end{table}

Changing only the material and the surface treatment on one and the same
structure varies $k_{\mathrm{eff}}$ by a factor of up to 23, and the
smaller the particle diameter, the larger the effect of the surface
treatment (\cref{fig:14}(a)).

Which branch type is limiting changes with the filler. For AlN most of the
heat passes through the contact branches, so the interfacial resistance in
series with them is limiting; for SiO$_2$ the near-contact branches through
the matrix carry more than a third of the total, so lowering the
interfacial resistance leaves $k_{\mathrm{eff}}$ almost unchanged
(\cref{fig:14}(b)).

The dissipation is concentrated on a small number of edges, and spreads out
on the side where the matrix is limiting. In a single run with $d=25\um$,
AlN and untreated surfaces, 63 edges -- the top 1\% of 6{,}289 -- carry
49\% of the total dissipation (\cref{fig:14}(c)).

The critical conductance level also shifts with the material. Comparing
conditions with the same particle diameter and the same treatment,
$\thcrit$ shifts by more than 1.5 decades with the filler alone, and the
shift due to the surface treatment is smaller than that
(\cref{fig:14}(d)). Because $\thcrit$ is obtained without solving a
Laplacian, it can be used for a first-pass screening of the material
choice.

The idealisation ratio $\Gamma_m$ of each mechanism (\cref{sec:4.5}) is
referred to each condition itself. For AlN, untreated, $d=5\um$ the three
mechanisms are comparable in magnitude, and applying the surface treatment
lowers the interface and raises the matrix. That a single change moves
which mechanism is limiting appears in the contrast between these two rows
(\cref{fig:14}(e)). For SiO$_2$ the gain from the interface
stays in the vicinity of unity, consistent with (b).

\section{Scope and limitations}\label{sec:7}

\subsection{Scope of the claims}\label{sec:7.1}

The claims of this paper are limited in four respects.

First, they are limited to the range of conditions used in the computation
(\cref{tab:1}), because none of the boundaries of the range of
applicability is available in the form of a threshold.

Second, the network of this paper carries edges only for particle pairs and
does not represent the parallel paths through the region of matrix far from
any particle. Unlike a mechanistic model that incorporates such a path into
the unit cell as an independent resistance~\citep{he2026}, ours
underestimates the effective thermal conductivity increasingly as the
packing fraction falls. On regular lattices the packing fraction below
which no edges are formed lies between 0.35 (simple cubic) and 0.49
(face-centred cubic), and the computation conditions of this paper for
random packings ($\phi\ge0.37$, \cref{tab:1}) lie above that lower bound.
The error estimate of the discrete network
approximation is given in terms of the perimeter of the regions containing
no particles~\citep{berlyand2012}, and the restriction of this subsection
states that condition from the side of the packing fraction.

Third, we have not cross-checked the absolute value of the effective
thermal conductivity against finite element analysis or against
measurement. What we have is agreement in the limits in which an exact
analytical solution is known, consistency with effective medium theory, and
the response to uncertainty in the material parameters (\cref{sec:5.3}).
The report that resistor network models in general agree with the finite
element method to within 4\% on average~\citep{birkholz2019} is a result
about the class of methods and not a validation of our model.

Fourth, we make no claim about rankings. The ranking of edges was given
in \cref{sec:4.3}. The same holds for the ranking of structures: for pairs
of structures whose effective conductances were matched by adjusting the
interfacial resistance, the critical conductance level hardly separated the
pairs at all, and we have confirmed that this is not the result of the
effects of structure and interfacial resistance cancelling (values in the
ancillary material, V15--V16). The only exactness this paper claims is that
of the sensitivity of \cref{sec:4}; the critical conductance level remains
a quantity that estimates the level of the effective conductance. The
critical path analysis put in correspondence in \cref{sec:3.6} is,
moreover, an argument that becomes asymptotically exact in the limit of an
infinitely broad conductance distribution~\citep{ahl1971}, and the
$\thcrit$ of the conditions of this paper (\cref{tab:8}) lies outside that
limit.

\subsection{Model assumptions and the limits of applicability}\label{sec:7.2}

The assumption that a particle is an isothermal node requires that the
through-particle resistance be small compared with the resistance between
particles. For a packing of equal spheres this ratio falls well below
unity and the assumption holds, and the resistance inside the particle
itself is included in the result as the through-particle resistance
(\cref{tab:2}). According to Batchelor and O'Brien, the isothermal
assumption fails for a pair with gap $h$ once $\Lambda=\alpha h/\rt$ falls
below unity~\citep{batchelor1977}; that effect is absorbed on the branch
side as the saturation shift of \cref{sec:2.2.2}. If a fibre is represented
as a chain of overlapping spheres, this ratio rises well above unity and
the scale of the contacts is undercounted as well, so for fillers of large
aspect ratio our model underestimates the thermal resistance (values in the
ancillary material, V17).

We do not introduce an interphase, that is, a coating layer of finite
thickness on the filler surface ($t_{\mathrm{ip}}=0$; the only exception is
the hypothetical case \texttt{lowk\_coat} of \cref{app:C}). Every real
surface treatment can be expressed as a difference in the interfacial
resistance $R_s$, whereas with a layer present the exchange between
branches becomes discontinuous at $h=2t_{\mathrm{ip}}$ and raising the
conductivity of the matrix and designing the coating layer can no longer be
decided independently.

The model is robust to non-uniformity of the surface treatment. Even when
30\% of the particles in one and the same structure are left untreated, the
ordering of the edges is largely preserved, and a surrogate model that
treats all particles as identically treated degrades the selection of the
top-ranked edges only slightly (values in the ancillary material, V18).

Beyond this, our model treats the steady state only and contains neither
heat capacity, nor temperature-dependent properties, nor many-body
screening. The descriptor returned by the filtration carries no information
about the direction of the heat flow either. Furthermore, in the stability
theorem of \cref{sec:3.5}, both a perturbation that moves the position of
the sink and a perturbation of the particle coordinates that changes the
edge set lie outside the premises. For a perturbation of the edge weights
alone the bound is attained almost with equality, whereas applying it
formally to a perturbation of the coordinates does not work as an estimate
(the procedure and the values are in the ancillary material, V19).

\subsection{Propagation of the uncertainty}\label{sec:7.3}

The material parameter with the widest uncertainty is the thermal contact
resistance $R_c$ between particles. As stated in \cref{sec:2.7}, there is
no established literature value, so for untreated particles we adopted
$2.5\times10^{-9}$ to $5\times10^{-8}\mKW$ as the assumed range. That
width is 1.30 decades. Computing one and the same structure at the two ends
of this range, the change in the edge conductances shrinks to a median of
0.212 decades for the edge that moves most, and to 0.102 decades in the
bottleneck distance between barcodes. The degree of the reduction,
$1.30/\|\Delta\log g\|_\infty$, depends on the ratio of the thermal
conductivities of filler and matrix, $\alpha=\kappa_f/\kmat$, and is larger
for smaller $\alpha$, as \cref{tab:9} shows. This is because for small
$\alpha$ the paths through the matrix dominate the resistance of the edge
and the contribution of the interface is relatively small. It follows that,
wherever the true value lies within the assumed range, the computation is
robust for fillers of low thermal conductivity. For fillers of high thermal
conductivity, on the other hand, the uncertainty in $R_c$ passes almost
unchanged into the uncertainty of the result, so measuring $R_c$ becomes an
open problem.

\begin{table}[tbp]
  \centering
  \caption{How the uncertainty in $R_c$ propagates (matrix: epoxy).}
  \label{tab:9}
  \begin{tabular}{@{}l r r@{}}
    \toprule
    Filler & $\alpha=\kappa_f/\kmat$
      & $1.30/\|\Delta\log g\|_\infty$ \\
    \midrule
    SiO$_2$      & 7       & 40.9 \\
    Al$_2$O$_3$  & 150     & 7.3 \\
    AlN          & 850     & 3.7 \\
    diamond      & 10{,}000 & 2.3 \\
    \bottomrule
  \end{tabular}
\end{table}

\section{Conclusion}\label{sec:8}

This paper has given a branch-conductance model that fixes the thermal
conductance of a particle pair (\cref{sec:2}), persistence relative to the
sink boundary on the axis of that conductance together with the critical
conductance level $\thcrit$ (\cref{sec:3}), and the exact equality of the
dissipation share with the logarithmic sensitivity of the effective
conductance together with the closed form for a finite change
(\cref{sec:4}). Whereas previous network descriptions remain at a
statistical relationship between an indicator of the structure and a
transport property, the difference here is that the allocation of the
dissipation among branch types is obtained as an exact sensitivity from one
sparse solve and one factorisation, and that the idealisation ratio of each
mechanism complements it as an exact value obtained by re-solving. What the
filtration returns, on the other hand, remains the level of the bottleneck
to percolation (\cref{sec:7.1}).

We list four open problems. First, to add the parallel paths through the
matrix to the edge set. It is the omission of those paths that fixes the
lower bound on the packing fraction (\cref{sec:7.1}). Second, to
cross-check the absolute value of the effective thermal conductivity
(\cref{sec:7.1}). Third, to explore whether any information beyond the
level can be extracted from the barcode, including statistics other than
$\thcrit$: for pairs matched in effective conductance the critical
conductance level hardly moves, whereas the gain ratio obtained from the
dissipation share spreads widely (\cref{sec:7.1}; values in the ancillary
material, V16). Fourth, to give a descriptor that distinguishes the
direction of the heat flow for anisotropic packed structures.

\appendix
\section{Derivations for the branch-conductance model}\label{app:A}

\subsection{The contact radius}\label{app:A.1}

When hard spheres of radii $r_i,r_j$ overlap by a depth $\delta$, the
centre-to-centre distance is $D=r_i+r_j-\delta$. The radius $a$ of the
circle in which the surfaces of the two spheres intersect is given, in
terms of the distance $x_i$ to the intersection plane, by
$x_i=(D^2+r_i^2-r_j^2)/(2D)$ and $a^2=r_i^2-x_i^2$. Splitting this as
$a^2=(r_i-x_i)(r_i+x_i)$ and putting each factor over the common
denominator $D$ gives
\begin{align}
  r_i-x_i&=\frac{r_j^2-(D-r_i)^2}{2D}=\frac{\delta\,(2r_j-\delta)}{2D},\\
  r_i+x_i&=\frac{(D+r_i)^2-r_j^2}{2D}
    =\frac{(2r_i-\delta)\big(2(r_i+r_j)-\delta\big)}{2D},
\end{align}
and multiplying them gives the closed form
\begin{equation}
  a^2=\frac{\delta\,(2r_i-\delta)(2r_j-\delta)
    \big(2(r_i+r_j)-\delta\big)}{4D^2}.
  \label{eq:a2closed}
\end{equation}
As $\delta\to0$ we have $D\to r_i+r_j$, so
\begin{equation}
  \frac{a^2}{\delta}\longrightarrow
  \frac{2r_i\cdot2r_j\cdot2(r_i+r_j)}{4(r_i+r_j)^2}=2\rt ,
\end{equation}
that is, $a\to\sqrt{2\rt\delta}$, which is $\sqrt2$ times the Hertz contact
radius $a_H=\sqrt{\rt\delta}$ (twice in area). Since our model generates
the structure from overlapping hard spheres, we use the intersection circle
as the contact face.

From the same expression, substituting $\delta=0$ directly gives
\begin{equation}
  x_i\big|_{\delta=0}
  =\frac{(r_i+r_j)^2+r_i^2-r_j^2}{2(r_i+r_j)}=r_i .
\end{equation}
This relation is the basis of the continuity of \cref{sec:2.6}.

\subsection{The annular integral of the near-contact branch}\label{app:A.2}

Under the gap $h(r)=h+r^2/(2\rt)$ of \cref{sec:2.2}, regarding each flux
tube of radius $r$ as passing in turn through interface, matrix and
interface and combining these in parallel in the radial direction gives
\begin{equation}
  g_{\mathrm{prox}}=\int_0^{r^{*}}
  \frac{2\pi r\,\mathrm dr}{(R_{s,i}+R_{s,j})+h(r)/k_{\mathrm{gap}}} .
  \label{eq:annular}
\end{equation}
This setting is of the same form as the analysis of Batchelor and
O'Brien~\citep{batchelor1977}. Substituting $u=r^2/(2\rt)$ gives
$r\,\mathrm dr=\rt\,\mathrm du$, and taking the outer cut-off at
$r^{*2}=2\rt^2$ makes the upper limit $\rt$. The antiderivative of the integrand is
$2\pi k_{\mathrm{gap}}\rt\,
\ln\big((R_{s,i}+R_{s,j})k_{\mathrm{gap}}+h+u\big)$, so
\begin{equation}
  g_{\mathrm{prox}}=\int_0^{\rt}
  \frac{2\pi\rt\,\mathrm du}{(R_{s,i}+R_{s,j})+(h+u)/k_{\mathrm{gap}}}
  =2\pi k_{\mathrm{gap}}\rt\,
  \ln\frac{(R_{s,i}+R_{s,j})k_{\mathrm{gap}}+h+\rt}
          {(R_{s,i}+R_{s,j})k_{\mathrm{gap}}+h} .
\end{equation}
Setting $a_{K,i}=R_{s,i}k_{\mathrm{gap}}$ gives $\ln\big(1+\rt/(h+a_{K,i}+a_{K,j})\big)$. The final form
at the head of \cref{sec:2.2} is this with the $h_{\mathrm{BOB}}$ of
\cref{sec:2.2.2} added to the denominator.

Because the interface and the matrix were combined in series within each
flux tube, $R_{s,i}+R_{s,j}$ appears at the same place in the integral as
$h/k_{\mathrm{gap}}$ and reduces to the substitution
$h\to h+a_{K,i}+a_{K,j}$. The order of combination cannot be reversed.
Combining layer by layer in the radial direction first makes the equivalent
heat transfer area of the gap layer,
$A=2\pi\rt h\ln(1+\rt/h)$, tend to zero with $h\to0$, so the series
interfacial term $(R_{s,i}+R_{s,j})/A$ diverges and $g\to0$, contradicting
the premise of the thin film approximation. The value of the outer cut-off
radius $r^{*}$ itself is not unique; the thin film approximation requires
only $r^{*}\gg\sqrt{\rt h}$.

\subsection{The matrix annulus of the contact branch}\label{app:A.3}

On a pair with $h\le0$, the matrix thickness at a radius $r>a$ is given by
the parabolic profile $h(r)=r^2/(2\rt)-\delta$, which is the integral of
\cref{app:A.2} with $h$ replaced by $-\delta$. The zero of this profile is
$r_0=\sqrt{2\rt\delta}$, which by the leading term of \cref{app:A.1}
differs from the exact intersection circle only at $O(\delta^2)$, so for
$\delta\ll\rt$ we may use $r_0$ as the lower limit of the integral. Using
the substitution of \cref{app:A.2} shifted by $\delta$,
$u=r^2/(2\rt)-\delta$, the lower limit $r=r_0$ moves to $u=0$ and the upper
limit $r=r^{*}$ to $u=\rt-\delta$, giving
\begin{equation}
  g_{\mathrm{ann}}=\int_{r_0}^{r^{*}}
  \frac{2\pi r\,\mathrm dr}{2R_s+h(r)/\kmat}
  =\int_0^{\rt-\delta}\frac{2\pi\rt\,\mathrm du}{2R_s+u/\kmat}
  =2\pi \kmat\rt\ln\!\Big(1+\frac{\rt-\delta}{a_{K,i}+a_{K,j}}\Big)
\end{equation}
The lower limit corresponds to $u=0$ after the substitution and therefore
does not appear
in the closed form, so the matrix annulus is fixed by the outer edge $\rt$
and the overlap $\delta$ alone (\cref{sec:2.3.2}).

The expression in \cref{sec:2.3.2} is this with the saturation shift of
\cref{sec:2.2.2} applied, that is, with $a_{K,i}+a_{K,j}$ replaced by
$a_{K,i}+a_{K,j}+h_{\mathrm{BOB}}$. The shift is introduced because the
logarithm above diverges in the limit $R_s=0$; it suppresses, with the same
value $h_{\mathrm{BOB}}=\rt/(\alpha^2-1)$, the same divergence as that of
the closed form of the near-contact branch as $h\to0$. Batchelor and
O'Brien give the flux through the contact circle as $2\kappa a$, from the
classical solution for a circular hole in a plate, and the flux through the
surrounding matrix as an integral with the prefactor
$2\pi\kmat\rt$~\citep{batchelor1977}. The constriction resistance of the
solid path and the matrix annulus are thus two terms of one and the same
analysis, with coefficients that are consistent with each other.

\subsection{The bond branch and discretisation}\label{app:A.4}

That the constriction resistance of the contact branch cannot be carried
over to the bond branch follows from the behaviour under discretisation.
The constriction resistance is the solution of the problem in which heat
enters a semi-infinite body from an isothermal disc of radius $a$, and it
does not depend on any interval length. Representing a solid of length
$\Lambda$ by $N$ spheres places $N$ such interval-independent quantities in
series, so the resistance of the whole chain grows in proportion to $N$ and
$g\to0$ as $N\to\infty$. With
$g_{\mathrm{bond}}=\kappa_h\pi R_{\min}^2/L$ of \cref{sec:2.4}, in
contrast, an $N$-fold subdivision makes each interval of length $L/N$, so
the resistance of the whole chain is unchanged.

\section{Proofs for the persistence and sensitivity results}\label{app:B}

We use the notation of \cref{sec:4.2}.

\subsection{Equivalent representations}\label{app:B.1}

\begin{figure}[tbp]
  \centering
  \includegraphics[width=\linewidth]{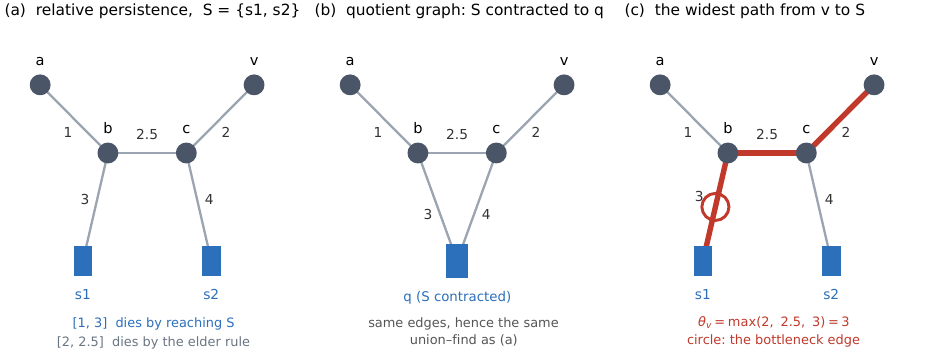}
  \caption{The three representations on a worked example: the relative
  barcode, the absolute barcode of the quotient graph, and the widest path
  values. The example includes a bar closed by the elder rule, whose death
  time is not a widest path value.}\label{fig:B1}
\end{figure}

We prove the two statements of \cref{sec:3.5}. First, the barcode of the
relative persistent homology $H_0(X_\theta,S)$ coincides with the barcode
of the absolute $H_0$ of the quotient graph in which $S$ is contracted to a
single point $q$. Second, the first $\theta$ at which the component
containing a vertex $v$ reaches $S$, which we write $\theta_v$, equals the
bottleneck value of the widest path from $v$ to $S$. The barcode also
contains bars that close by merging into an older component without
reaching $S$ (the elder rule), and their death times are not widest path
values; the second statement concerns the death times of the bars closed by
reaching $S$. The correspondence between the three representations is
illustrated on the example of \cref{fig:B1}.

\begin{proof}
The first coincidence follows from the construction. Contracting collects
the interior of $S$ into a single point and changes neither the vertices
nor the edges outside $S$. The union--find process that adds edges in
increasing order of $\theta$ is therefore identical in the two cases, and
the birth of a component, a merger, and the connection to $S$ (to $q$ after
contraction) all occur at the same times.

We now prove the second statement. Since $\theta=-\log_{10}g$, increasing
order in $\theta$ is decreasing order in conductance. By the definition of
$\theta_v$ there is a path from $v$ to $S$ using only edges with
$\theta_e\le\theta_v$. The maximum of $\theta_e$ along this path is at most
$\theta_v$, so the value obtained by minimising that maximum over all paths
-- the bottleneck value of the widest path, which in terms of conductance
is the maximised minimum $10^{-\theta_v}$ -- is at most $\theta_v$.
Conversely, if that minimised value were smaller than $\theta_v$, there
would exist a path all of whose edges are added before $\theta_v$, so the
component of $v$ would reach $S$ before $\theta_v$, contradicting the
definition of $\theta_v$. The two are therefore equal.
\end{proof}

\subsection{The envelope theorem and the logarithmic sensitivity}\label{app:B.2}

At the minimiser,
\begin{equation}
  \frac{\partial \Geff}{\partial g_e}=(\mathrm du^\star)_e^2 .
\end{equation}

\begin{proof}
Write the Dirichlet energy as $F(u,g)=\sum_{e}g_e(\mathrm du)_e^2$. Let
$\Delta>0$ and $g'=g+\Delta\mathbf e_e$. Since $\Geff$ is a minimum, using
the minimiser $u^\star_g$ of $g$ as a trial function for $g'$ gives
\begin{equation}
  \Geff(g')\le F(u^\star_g,g')=\Geff(g)+\Delta\,(\mathrm du^\star_g)_e^2,
\end{equation}
and conversely, using the minimiser $u^\star_{g'}$ of $g'$ as a trial
function for $g$ gives
$\Geff(g)\le \Geff(g')-\Delta\,(\mathrm du^\star_{g'})_e^2$. Together,
\begin{equation}
  (\mathrm du^\star_{g'})_e^2\;\le\;\frac{\Geff(g')-\Geff(g)}{\Delta}
  \;\le\;(\mathrm du^\star_g)_e^2
\end{equation}
(the inequalities reverse for $\Delta<0$), and by the continuity of
$(\mathrm du^\star)_e$ in $g$ the two sides coincide as $\Delta\to0$, which
gives the statement. That $(\mathrm du^\star)_e$ is unique and continuous
follows because $F$ is a convex quadratic form in the interior variables
and the difference of any two minimisers is constant on each connected
component.
\end{proof}

This estimate uses only that $\Geff$ is defined as a minimum and that the
boundary values $1$ and $0$ do not depend on $g$ (the setting of
\cref{sec:2.1}). Differentiating as a composite function,
\begin{equation}
  \frac{\mathrm d}{\mathrm dg_e}F(u^\star(g),g)
  =\underbrace{\frac{\partial F}{\partial g_e}}_{=(\mathrm du^\star)_e^2}
  +\sum_{v}\frac{\partial F}{\partial u_v}\bigg|_{u^\star}
  \frac{\partial u^\star_v}{\partial g_e},
\end{equation}
and the result above amounts to the vanishing of the second term. At
interior vertices the gradient of $F$ is zero, that is, the current law
holds at every vertex, and on the boundary $u^\star_v$ does not depend on
$g$. On components not connected to the boundary a constant per component
remains free in $u^\star$, but $(\mathrm du^\star)_e$ is uniquely zero, so
the identity holds. The form in which the derivative of a minimum with
respect to a parameter equals the partial derivative at a fixed minimiser
is the classical case of the envelope theorem, and for networks it is known
as sensitivity analysis by the adjoint network method.

Combining this with the sum rule gives the logarithmic sensitivity. Because
the boundary conditions are normalised to $1$ and $0$,
\begin{equation}
  \sum_{e\in E}P_e=\sum_{e\in E}g_e(\mathrm du^\star)_e^2=F(u^\star,g)=\Geff
\end{equation}
follows from the definitions, and therefore
\begin{equation}
  \frac{\partial\log \Geff}{\partial\log g_e}
  =\frac{g_e}{\Geff}\frac{\partial \Geff}{\partial g_e}
  =\frac{g_e(\mathrm du^\star)_e^2}{\Geff}=\Pshare_e .
\end{equation}

\subsection{The closed form for a finite change}\label{app:B.3}

For an edge $e=(i,j)$ we establish the closed form of \cref{sec:4.4},
\begin{equation}
  \Geff(g+\Delta\mathbf e_e)=\Geff(g)
  +\frac{\Delta\,(\mathrm du^\star)_e^2}{1+\Delta\,\Rt_e},
\end{equation}
where $\Rt_e$ is the effective resistance between $i$ and $j$ in the
circuit in which the source and the sink are shorted to a single point.

\begin{proof}
The change of the Laplacian is of rank one: with $b=\mathbf e_i-\mathbf e_j$
we may write $L(g+\Delta\mathbf e_e)=L(g)+\Delta\,bb^{\top}$. Applying the
Sherman--Morrison formula to the Schur complement under the boundary
conditions gives the expression above. A formula of the same form is given
as a rank-one perturbation of the internal network of a Kron
reduction~\citep{dorfler2013}, and the quantity appearing in its
denominator carries the same interpretation, as the effective resistance of
the internal network defined by the submatrix with the boundary nodes
removed.
\end{proof}

Furthermore,
\begin{equation}
  \frac{\mathrm d\Geff}{\mathrm d\Delta}
  =\frac{(\mathrm du^\star)_e^2}{(1+\Delta\Rt_e)^2}>0
\end{equation}
is decreasing in $\Delta$, so $\Geff$ is monotonically increasing and
concave in each $g_e$.

\subsection{Contrast with an existing sensitivity indicator}\label{app:B.4}

The weight $g_eR_{\mathrm{eff}}(e)$ by effective resistance of
\cref{sec:4.6}~\citep{spielman2011}, in which $R_{\mathrm{eff}}$ is the
ordinary effective resistance without shorting, can also be written as a
logarithmic derivative, but of a different object. Writing $T(G)$ for the
weighted number of spanning trees,
\begin{equation}
  g_eR_{\mathrm{eff}}(e)=\frac{\partial\log T(G)}{\partial\log g_e} .
\end{equation}
This follows because $T(G)$ equals a cofactor of the Laplacian by the
matrix-tree theorem, and because the logarithmic derivative of a
determinant can be written as a product with the inverse matrix.

The normalisation differs as well. The identity
\begin{equation}
  \sum_e g_eR_{\mathrm{eff}}(e)=n-1
\end{equation}
follows because $W^{1/2}BL^{+}B^{\mathsf T}W^{1/2}$ is a projection of rank
$n-1$~\citep{spielman2011}, and it is fixed by the size of the graph alone,
independently of the edge weights. By \cref{app:B.2}, in contrast,
$\sum_e\Pshare_e=1$.

\section{Structure families}\label{app:C}

The generation rules for the 23 structure families that make up the
ensemble of this paper are given in \cref{tab:C1} and the generated
properties in \cref{tab:C2}. Generation uses
\texttt{ensemble.make(family, d\_um, seed)} of the accompanying software
package, and the values in the tables are medians over three random seeds.
The regression, the cross-validation and the residual (0.250 decades) of
this paper all use the 69 packings formed from the 23 families with three
seeds each. For the sedimentation, $s=0$ is none and $s=1$ is complete,
and stratification is given by $z'=L(z/L)^{1+p}$.

\begin{table}[tbp]
  \centering
  \caption{Generation rules of the structure families.}\label{tab:C1}
  \small
  \setlength{\tabcolsep}{4pt}
  \begin{tabular}{@{}p{0.17\linewidth} p{0.19\linewidth} p{0.55\linewidth}@{}}
    \toprule
    Family & Group & Generation rule (difference from the default) \\
    \midrule
    \texttt{phi\_050} & reference series & none (isotropic random packing) \\
    \texttt{phi\_055} & reference series & none (isotropic random packing) \\
    \texttt{phi\_060} & reference series & none (isotropic random packing) \\
    \texttt{phi\_064} & reference series & none (isotropic random packing) \\
    \texttt{bimodal\_3to1} & high loading & bimodal, size ratio 3, volume
      fraction of the large particles 0.7 \\
    \texttt{bimodal\_5to1} & high loading & bimodal, size ratio 5, volume
      fraction of the large particles 0.75 \\
    \texttt{poly\_wide} & high loading & continuous distribution,
      $\sigma_{\ln}=0.5$ (lognormal width) \\
    \texttt{voided} & poor dispersion & spherical void, radius $0.26L$ \\
    \texttt{clustered} & poor dispersion & agglomeration, strength 0.32 \\
    \texttt{stratified} & process segregation & stratification $p=0.55$ \\
    \texttt{segregated} & process segregation & bimodal, size ratio 3,
      volume fraction of the large particles 0.7; sedimentation $s=1$ \\
    \texttt{fiber\_z} & anisotropy & fibres oriented along $z$,
      24 fibres $\times$ 6 spheres \\
    \texttt{fiber\_xy} & anisotropy & fibres oriented in $xy$,
      24 fibres $\times$ 6 spheres \\
    \texttt{coat\_mix} & non-uniform treatment & 30\% low-$\kappa$ coated
      particles (\texttt{lowk\_coat}) \\
    \texttt{strat\_025} & segregation strength & stratification $p=0.25$ \\
    \texttt{strat\_055} & segregation strength & stratification $p=0.55$ \\
    \texttt{strat\_100} & segregation strength & stratification $p=1$ \\
    \texttt{strat\_180} & segregation strength & stratification $p=1.8$ \\
    \texttt{seg\_000} & segregation strength & bimodal, size ratio 3,
      volume fraction of the large particles 0.7; sedimentation $s=0$ \\
    \texttt{seg\_033} & segregation strength & as above,
      sedimentation $s=0.33$ \\
    \texttt{seg\_067} & segregation strength & as above,
      sedimentation $s=0.67$ \\
    \texttt{seg\_100} & segregation strength & as above,
      sedimentation $s=1$ \\
    \texttt{jittered\_ref} & legacy baseline & jittered lattice \\
    \bottomrule
  \end{tabular}
\end{table}

\begin{table}[tbp]
  \centering
  \caption{Generated properties of the structure families (medians over
  three seeds; the packing generator, version 4).}\label{tab:C2}
  \footnotesize\setlength{\tabcolsep}{3pt}
  \begin{tabular}{@{}l r r c r r c r@{}}
    \toprule
    Family & $\phi$ & $N$ & $d$ [$\mu$m] & Ratio & CV & Box [$\mu$m]
      & $L/A$ [1/$\mu$m] \\
    \midrule
    \texttt{phi\_050} & 0.5000 & 900 & 20.0--20.0 & 1.00 & 0.000
      & $196^3$ & 0.0051 \\
    \texttt{phi\_055} & 0.5500 & 900 & 20.0--20.0 & 1.00 & 0.000
      & $190^3$ & 0.0053 \\
    \texttt{phi\_060} & 0.6000 & 900 & 20.0--20.0 & 1.00 & 0.000
      & $185^3$ & 0.0054 \\
    \texttt{phi\_064} & 0.6400 & 900 & 20.0--20.0 & 1.00 & 0.000
      & $181^3$ & 0.0055 \\
    \texttt{bimodal\_3to1} & 0.6600 & 900 & 13.7--41.2 & 3.00 & 0.468
      & $179^3$ & 0.0056 \\
    \texttt{bimodal\_5to1} & 0.7000 & 900 & 12.7--63.6 & 5.00 & 0.552
      & $175^3$ & 0.0057 \\
    \texttt{poly\_wide} & 0.6400 & 900 & 3.0--67.5 & 22.54 & 0.538
      & $181^3$ & 0.0055 \\
    \texttt{voided} & 0.6000 & 900 & 20.0--20.0 & 1.00 & 0.000
      & $185^3$ & 0.0054 \\
    \texttt{clustered} & 0.6000 & 900 & 20.0--20.0 & 1.00 & 0.000
      & $185^3$ & 0.0054 \\
    \texttt{stratified} & 0.6000 & 900 & 20.0--20.0 & 1.00 & 0.000
      & $185^3$ & 0.0054 \\
    \texttt{segregated} & 0.6600 & 900 & 13.7--41.2 & 3.00 & 0.468
      & $179^3$ & 0.0056 \\
    \texttt{fiber\_z} & 0.6960 & 1044 & 20.0--20.0 & 1.00 & 0.000
      & $185^3$ & 0.0054 \\
    \texttt{fiber\_xy} & 0.6960 & 1044 & 20.0--20.0 & 1.00 & 0.000
      & $185^3$ & 0.0054 \\
    \texttt{coat\_mix} & 0.6000 & 900 & 20.0--20.0 & 1.00 & 0.000
      & $185^3$ & 0.0054 \\
    \texttt{strat\_025} & 0.6000 & 900 & 20.0--20.0 & 1.00 & 0.000
      & $185^3$ & 0.0054 \\
    \texttt{strat\_055} & 0.6000 & 900 & 20.0--20.0 & 1.00 & 0.000
      & $185^3$ & 0.0054 \\
    \texttt{strat\_100} & 0.6000 & 900 & 20.0--20.0 & 1.00 & 0.000
      & $185^3$ & 0.0054 \\
    \texttt{strat\_180} & 0.6000 & 900 & 20.0--20.0 & 1.00 & 0.000
      & $185^3$ & 0.0054 \\
    \texttt{seg\_000} & 0.6600 & 900 & 13.7--41.2 & 3.00 & 0.468
      & $179^3$ & 0.0056 \\
    \texttt{seg\_033} & 0.6600 & 900 & 13.7--41.2 & 3.00 & 0.468
      & $179^3$ & 0.0056 \\
    \texttt{seg\_067} & 0.6600 & 900 & 13.7--41.2 & 3.00 & 0.468
      & $179^3$ & 0.0056 \\
    \texttt{seg\_100} & 0.6600 & 900 & 13.7--41.2 & 3.00 & 0.468
      & $179^3$ & 0.0056 \\
    \texttt{jittered\_ref} & 0.3737 & 1036 & 20.0--25.0 & 1.25 & 0.020
      & $355\times93\times355$ & 0.0108 \\
    \bottomrule
  \end{tabular}
\end{table}

The packing fraction is generated as targeted. The exceptions are
\texttt{fiber\_z} and \texttt{fiber\_xy}, and \texttt{jittered\_ref}, for
which no target is set. The former two are made by adding fibres (24 fibres
of 6 spheres) to an isotropic packing at a packing fraction of 0.600, so
the packing fraction after generation is 0.696 and the number of particles
is 1{,}044.

Four points should be noted in reading the tables. First, the size ratio of
22.5 for
\texttt{poly\_wide} is a value that follows from the width of the
distribution ($\sigma_{\ln}=0.5$), and differs in character from the 3.00
and 5.00 of the bimodal families, which are design values. Second, the
dimensions of the simulation box (the specimen) differ by family, and only
\texttt{jittered\_ref} is not a cube ($355\times93\times355\um$). Third,
\texttt{jittered\_ref} is a legacy family generated by a different method,
but it is not excluded and is used as part of the 69 packings. Fourth, the
coated particles of \texttt{coat\_mix} are represented by the treatment
\texttt{lowk\_coat} ($t_{\mathrm{ip}}=0.05\um$,
$k_{\mathrm{ip}}=0.05\WmK$, $R_c=1\times10^{-6}\mKW$). It is a
hypothetical case representing the extreme on the unfavourable side rather
than a real material, and it is the only exception to $t_{\mathrm{ip}}=0$
in \cref{tab:1}.

\bibliographystyle{unsrtnat}
\bibliography{refs}

\end{document}